\documentstyle[prl,aps,epsfig]{revtex}

\def\Journal#1#2#3#4{{#4} {#1} {\bf #2}  #3}
\def\CJF{{\em Czech. J. Phys} B}
\def\ARNPS{\em Ann. Rev. Nucl. Part. Sci.}
\def\IZV{\em Izv. AN SSSR}
\def\JINR{\em JINR Communications}
\def\PPNP{\em Prog. Part. Nucl. Phys.}
\def\PTPS{\em Prog. Theor. Phys. (Supp.) }
\def\PTP{\em Prog. Theor. Phys. }
\def\NCA{\em Nuovo Cimento}
\def\JETL{\em JETP Lett.}
\def\FF{\em Found. Phys.}

\def\NPB{{\em Nucl. Phys.} B}
\def\NPBP{{\em Nucl. Phys.} B{(Proc. Suppl.)}}
\def\NP{{\em Nucl. Phys.} }
\def\NPA{{\em Nucl. Phys.} A}
\def\PLB{{\em Phys. Lett.}  B}
\def\PRL{\em Phys. Rev. Lett.}
\def\ZETF{\em Zh. Eksp. Theor. Fiz.}
\def\PRD{{\em Phys. Rev.} D}
\def\PR{{\em Phys. Rev.}}
\def\PRC{{\em Phys. Rev.} C}
\def\PRP{\em Phys. Rep.} 
\def\MPL{{\em Mod. Phys. Lett.} A} 
\def\ZPC{{\em Z. Phys.} C}
\def\ZPA{{\em Z. Phys.} A}
\def\JPG{{\em J. Phys.} G}
\def\RPP{\em Rep.Prog. Phys.} 
\def\RMP{\em Rev. Mod. Phys.} 
\def\SPU{\em Sov. Phys. Usp.} 
\def\SJNP{\em Sov. J. Nucl. Phys.} 
\def\FIZB{{\em FIZIKA} B} 

\def\be{\begin{equation}}
\def\ee{\end{equation}}
\def\bea{\begin{eqnarray}}
\def\eea{\end{eqnarray}}

\begin{document}
\draft                                                            
\title{Double beta decay\footnote{Supported by the Deutsche
Forschungsgemeinschaft Fa67/17 and Fa67/21 and the EU under contract CT94-0603
and CT93-0323}}
\author{Amand Faessler$^1$ and Fedor \v Simkovic$^2$}

\address{1.  Institute f\"ur Theoretische Physik der Universit\"at 
T\"ubingen\\ Auf der Morgenstelle 14, D-72076 T\"ubingen, Germany \\
e-mail: amand.faessler@uni-tuebingen.de\\
2. Department of Nuclear Physics,  Comenius University,\\
Mlynsk\'a dol., pav. F1, SK-842 15 Bratislava, Slovakia\\  
e-mail: simkovic@fmph.uniba.sk }

\date{\today}
\maketitle
\begin{abstract}
We review the recent developments in the field of 
nuclear double beta decay, which is presently an  important topic 
in both nuclear and particle physics. 
The mechanism of lepton number violation within the neutrinoless
double beta decay ($0\nu\beta\beta$-decay) is discussed in context
of the problem of neutrino mixing  and the R-parity violating 
supersymmetric extensions of the Standard model.   
The problem of reliable determination of the nuclear
matrix elements governing both two-neutrino and neutrinoless
modes of the double beta decay is addressed. 
The validity of different approximation schemes in the 
considered nuclear structure studies is analyzed and 
the role of the Pauli exclusion principle for a correct treatment 
of nuclear matrix elements is emphasized. 
The constraints on different lepton number violating parameters
like effective electron 
neutrino mass, effective right-handed weak interaction
parameters, effective Majoron coupling constant and 
R-parity violating SUSY parameters are derived from the best presently
available experimental limits on the half life of 
$0\nu\beta\beta$-decay. 
\end{abstract}
\widetext

\tableofcontents

\section{Introduction}

The double beta ($\beta\beta$) decay  attracts the attention of both 
experimentalists and theoreticians for already a long period and 
remains of major importance both for particle and nuclear physics. 

The double beta decay is a second order process of weak interaction
and there are few tenths of nuclear systems \cite{hax84}, which offer 
an opportunity to study it. The $\beta\beta$ decay can be observed
because the pairing force 
renders the even-even nuclei with even number of protons and
neutrons more stable than the odd-odd nuclei with broken pairs. Thus, 
the single beta decay transition from the even-even parent nucleus 
(A,Z) to the neighboring odd-odd nucleus (A,Z+1) is forbidden 
energetically and the $\beta\beta$ decay to the daughter nucleus 
(A,Z+2) is the only possible decay channel. 

There are different possible modes of the double beta decay, which
differ from each other by the light particles accompanying 
the emission of two electrons. We distinguish the double beta decay 
modes with and without lepton number violation.

The two-neutrino
double beta decay ($2\nu\beta\beta$-decay), which involves the 
emission of two electrons and two antineutrinos,
\begin{equation}
(A,Z) \rightarrow (A,Z+2) + 2e^- +2{\overline{\nu}}_e,
\label{int.1}
\end{equation}
is a process fully consistent with the standard model (SM) of
electroweak interaction formulated by Glashow \cite{gla61}, 
Weinberg \cite{wei67} 
and Salam \cite{sal68}. This decay mode with obvious lepton number
conservation was first considered by Mayer in 1935\cite{may35}. 
The inverse half-life of $2\nu\beta\beta$-decay is free of 
unknown parameters on the particle physics side and is expressed as 
a product of  a phase-space factor and the relevant 
$2\nu\beta\beta$-decay nuclear matrix element. Thus, the measured 
experimental half lifes of 
$2\nu\beta\beta$-decays give us directly the value of the 
$2\nu\beta\beta$-decay 
nuclear matrix elements. In this way $2\nu\beta\beta$-decay offers a
severe test of nuclear structure calculations.

The $2\nu\beta\beta$-decay is already well established experimentally 
for a couple of isotopes. The most favored for the experimental study
of this rare process
is the transition from the ground state $0^+$ of the initial  
to the ground state $0^+$ of the final nuclei because of the 
large energy release. But recently, increased attention is paid
also for transitions to the $2^+$ and $0^+$ excited states of the final
nucleus 
\cite{barb90,kopy90,blum92,piep94,barb95,suc93,suc94,sto94,dhi95,sch97,bara97}.
The half-lifes of the $2\nu\beta\beta$-decay range 
from $10^{19}$ up to $10^{24}$ years 
[see Table \ref{tabint.1}]. Direct counter 
experiments have observed  $2\nu\beta\beta$-decays for  ground state to
ground state transitions in $^{48}Ca$,  $^{76}Ge$,
$^{82}Se$, $^{100}Mo$, $^{116}Cd$ and $^{150}Nd$. A positive evidence 
for a $2\nu\beta\beta$-decay transition to the $0^+_1$ excited state of 
the final
nucleus was observed for $^{100}Mo$ \cite{barb95}. The
geochemical experiments, which observe the double beta decay through
daughter isotope excesses in natural samples
has proved the existence of double beta decay in $^{82}Se$,
$^{96}Zr$, $^{128}Te$ and $^{130}Te$. The radiochemical experiment,
which observes the accumulation of daughter isotopes under laboratory
condition, has observed double beta decay in  $^{238}U$.
The positive signals of geochemical and radiochemical experiments
are identified with the lepton number conserving 
$2\nu\beta\beta$-decay mode.

The neutrinoless mode of the double beta decay ($0\nu\beta\beta$-decay)
(proposed by Fury in 1939 \cite{fur39}), 
which involves the emission of two electrons and no neutrinos 
\begin{equation}
(A,Z) \rightarrow (A,Z+2) + 2e^-,
\label{int.2}
\end{equation}
is expected to occur if lepton number conservation is not an exact symmetry
of nature and thus is forbidden in the SM (Standard Model) of 
electroweak interaction.
The $0\nu\beta\beta$-decay takes place only if the neutrino is a
Majorana particle (i.e. identical to its own antiparticle) with non-zero 
mass.

The study of the $0\nu\beta\beta$-decay 
is stimulated by the development of grand
unified theories (GUT's) \cite{moh91,val92}
representing generalization the 
of $SU(2)_L \otimes U(1)$ SM. In spite of 
the facts that the SM represents the simplest and most economical theory,
which describes jointly weak and 
electromagnetic interactions and in spite of the fact 
that it has been very successful, wherever it has been tested, the SM can 
not answer many of the fundamental questions, e.g.: 
Is the neutrino really massless? If it has mass,
why is this mass much smaller than that of corresponding charged leptons?
What kind of particles are neutrinos, Dirac or Majorana? Does neutrino
mixing take place? Therefore, the SM cannot be considered as the ultimative 
theory of nature. 
The non-zero neutrino masses and neutrino mixing appear naturally in many
different variants of GUT's like the simplest
SO(10) left-right symmetric model \cite{moh92}, 
minimal supersymmetric standard
model (MSSM) and their extensions. The exception is essentially the minimal
SU(5) model \cite{geo74}, which however has been practically ruled out due 
unsuccessful searches for nucleon instability.
The expectations arising from GUT's 
are that the conservation laws of the SM may be violated 
to some small degree. The GUT's offer a variety of mechanisms
which allow the $0\nu\beta\beta$-decay, from which the well-known
possibility is via the exchange of a Majorana neutrino between the two
decaying neutrons \cite{hax84,sch84,doi85,ver86}. 

If the global symmetry 
associated with lepton number conservation is broken spontaneously, 
the models imply the existence of a physical Nambu-Goldstone boson, 
called Majoron \cite{chi81,gel81,geo81,sch82}, which couples to
neutrinos. The Majoron might occur in the Majoron mode of the 
neutrinoless double beta decay 
\cite{ber92,hir96}:
\begin{equation}
(A,Z) \rightarrow (A,Z+2) + 2e^- + \phi.
\label{int.3}
\end{equation}
There are also other possible mechanisms of $0\nu\beta\beta$-decay
induced by lepton-number violating quark-lepton interactions
of R-parity non-conserving extensions 
of the SM \cite{moh86,ver87}. A complete analysis of this mechanism 
within the MSSM for the case the initial d-quarks are put inside the
two initial neutrons (two-nucleon SUSY mode) was carried in \cite{hkk96}. 
Recently, it has been found that a new contribution of the 
R-parity violating ($R_p \hspace{-1em}/\;\:$) supersymmetry (SUSY) to the 
$0\nu\beta\beta$ via pion exchange between the two decaying neutrons
dominates over the two-nucleon $R_p \hspace{-1em}/\;\:$  
SUSY mode \cite{fae97}.  The R-parity conserving SUSY mechanisms of 
$0\nu\beta\beta$-decay have been proposed and investigated in 
Ref. \cite{hir97}. The GUT's predict also new type of gauge bosons
called leptoquarks,
which can transform quarks into leptons or vice versa \cite{buc87}.
A new mechanism for $0\nu\beta\beta$-decay based on leptoquark 
exchange has been discussed in \cite{lpq96}.

The $0\nu\beta\beta$-decay has not been seen experimentally till now. 
We note that it is easy to distinguish between the three decay modes 
in Eqs. (\ref{int.1})-(\ref{int.3}) in the experiment 
by measuring the sum of electron energies \cite{hax84}. A signal from
the $0\nu\beta\beta$-decay is expected to be a peak at the end of 
the electron-electron coincidence spectrums as a function of the sum 
of the energies of the two electrons  
as they carry the full 
available kinetic energy for this process. Both, 
the $2\nu\beta\beta$-decay and the $0\nu\beta\beta\phi$-decay modes yield
a continuous electron spectrum, which differ by the position of the maximum
as different numbers of light particles are present in the final state. 
The observation of $0\nu\beta\beta$-decay would signal physics beyond the SM. 
It is worthwhile to notice that the $0\nu\beta\beta$-decay is one of
few eminent non-accelerator experiments which may probe grand 
unification scales far beyond present and future accelerator energies. 
For these reasons, considerable experimental effort is being
devoted to the study of this decay mode. Presently there
are about 40 experiments
searching for  the $0\nu\beta\beta$-decay. The best available 
experimental limits on the half-life of this decay mode 
for different nuclei are presented in Table \ref{tabint.1}. 
The most stringent 
experimental lower bound for a half-life has been measured by the 
Heidelberg-Moscow collaboration for the $0\nu\beta\beta$-decay of $^{76}Ge$
\cite{bau97}:
\begin{eqnarray}
T_{1/2}^{{0\nu\beta\beta}-\mbox{exp}}(0^+ \rightarrow 0^+)
\hskip2mm \geq \hskip2mm
1.1 \times 10^{25} \mbox{ years} \ \ \ \ \ \ \ \ (90 \% \ \mbox{C.L.})
\end{eqnarray}
The experimental half-life 
limits allow to extract limits on different lepton number 
violating parameters, e.g. effective neutrino mass, 
parameters of right-handed 
currents and  parameters of supersymmetric models.

However, in order to correctly interpret the results of 
$0\nu\beta\beta$-decay
experiments, i.e. to obtain quantitative answers for the lepton number 
violating parameters, the mechanism of nuclear transitions has to be 
understood. This means one has to evaluate
the corresponding nuclear matrix elements 
with high reliability.  

The calculation of the nuclear many-body Green 
function governing the double beta decay transitions 
continues to be challenging and attracts the specialists of different
nuclear models. As  each mode of double beta decay requires 
the construction of the same many-body nuclear structure wave functions,
the usual strategy has been first to try to reproduce the 
observed $2\nu\beta\beta$-decay half-lifes.  A comparison 
between theory and experiment for the $2\nu\beta\beta$ - decay 
provides a measure of confidence in the calculated nuclear wave 
functions employed for extracting the unknown parameters from 
$0\nu\beta\beta$-decay life measurements.

A variety of nuclear techniques have been used in attempts 
to calculate the nuclear 
$2\nu\beta\beta$-decay matrix elements, which 
require the summation over a complete set of intermediate
nuclear states. It has been found and it is widely accepted 
that the $2\nu\beta\beta$-decay matrix elements are strongly 
suppressed. The nuclear systems which can undergo double 
beta decay are medium and heavy open-shell nuclei with a complicated 
nuclear structure. The only exception is the A=48 system.  
Within the shell model, which describes well the low-lying states in the 
initial and final nuclei, it is clearly impossible to construct 
all the needed states of the intermediate nucleus. Therefore,
the proton-neutron Quasiparticle Random Phase Approximation (pn-QRPA),
which is one of the approximation to the many-body problem,
has developed into one of the most popular methods for calculating
nuclear wave functions involved in the $\beta\beta$  decay 
\cite{vog86,civ87,mut88}.
The QRPA plays a prominent role in the analysis, which are 
unaccessible to shell model calculations. The QRPA has been found
successful in revealing the suppression mechanism 
for the $2\nu\beta\beta$-decay
\cite{vog86,civ87,mut88}.
However, the predictive power of the QRPA is questionable because of
the extreme sensitivity of the calculated $2\nu\beta\beta$-decay 
matrix elements in the physically acceptable region on the 
particle-particle strength of the nuclear Hamiltonian.

This quenching behavior of the $2\nu\beta\beta$-decay matrix elements is a
puzzle and has attracted the attention of many theoreticians.
The main drawback of the QRPA is the overestimation of the ground state 
correlations leading to a collapse of the QRPA ground state. 
Several attempts have been made in the past to shift the collapse of the 
QRPA to higher values by proposing  different alterations of the QRPA
including proton-neutron pairing \cite{cheo93}, higher order RPA
corrections \cite{rad91} and particle number projection \cite{civ91,krm93}.
But, none of them succeeded to avoid collapse of the QRPA and to reduce 
significantly the dependence of $2\nu\beta\beta$-decay 
matrix elements on the particle-particle strength.  It is because 
all these methods disregard the main source of the ground state
instability which is traced to the assumption of the 
quasiboson approximation (QBA) violating the Pauli exclusion principle.  
Recently, Toivanen and Suhonen have proposed a proton-neutron
renormalized QRPA (pn-RQRPA) to study the two neutrino \cite{toi95} and
Schwieger, Simkovic and Faessler the $0\nu \beta \beta$ decay \cite{simn96}. 
In these works 
the Pauli exclusion principle is taken into account more carefully.
The pn-RQRPA prevents to build too strong ground state correlations
and avoids the collapse of the pn-RQRPA solution that plaque the pn-QRPA.
The $2\nu\beta\beta$-decay 
matrix elements calculated within pn-RQRPA and pn-RQRPA with 
proton-neutron pairing (full-RQRPA) \cite{simn96} has been found 
significantly less sensitive to the increasing strength of the 
particle-particle interaction in comparison with pn-QRPA  results.  
In the meanwhile, some critical studies have discussed the 
shortcomings of RQRPA  like the violation of the  Ikeda sum rule 
\cite{krm96,hirm96} and particle number non-conservation 
in average in respect
to the  correlated ground states in the double beta decay 
\cite{simm97,sims97}.

A large amount of theoretical work has been done to calculate nuclear 
matrix elements
and decay rates. However, the mechanism which is leading to the suppression 
of these matrix elements is still not completely understood. 
The practical calculation always involves some approximations, which
make it difficult to obtain an unambiguous decay rate.
The calculation of the  $2\nu\beta\beta$-decay nuclear transition 
continues to be subject of interest,  which stimulates the rapid 
development of different approaches to the many-body problem, e.g.
shell model and QRPA.

The aim of this review is to present some of the most 
recent developments in the field of the double beta decay.
Let us note that several other review articles treating the 
problem of double beta decay already exist. 
The theory and the experimental status of double beta decay 
have been reviewed e.g. by Primakoff and Rosen (1981) \cite{pri81},
Boehm and Vogel (1984) \cite{boe84}, 
Haxton and Stephenson (1984) \cite{hax84}, 
Schepkin (1984) \cite{sch84}, Doi et al (1985) \cite{doi85}, 
Vergados (1986) \cite{ver86}, Avignone and Brodzinski (1988) \cite{avi88},
Grotz and Klapdor \cite{gro90}, Tomoda (1991) \cite{tom91}, 
and Moe and Vogel (1994) \cite{moe94}.

\section{Massive neutrinos}

The neutrino remains a puzzle. The neutrino is the only elementary particle,
which basic properties are not known till today. 
In contrast with the charged fermions the nature and the masses 
of the neutrinos has not yet been established phenomenologically. 
The neutrino can be like other fermions
a Dirac particle, i.e. is different from its antiparticle. However, 
there is an another possibility proposed long ago by Majorana \cite{maj37}. 
The neutrino is the only fermion which can be a Majorana particle, i.e. 
it is identical to its antiparticle. 
This distinction is important if the mass of
neutrino is non-zero.  It is worthwhile to notice that there is no principle,
which dictates neutrinos to be massless 
particles as it is postulated in the SM. 
If the neutrinos are  particles with definite mass, it is natural to suppose
that neutrino mixing does take place. This was first considered by 
Pontecorvo \cite{pon57,pon58}. It means that flavor neutrino
fields in the weak interaction Lagrangian
are a mixture of fields of neutrinos possessing definite masses.  
Unlike of only one possible scheme for quark mixing there exist 
several totally different schemes of neutrinos mixing as the 
neutrino is electrically neutral.  The number of massive 
neutrinos depends on the considered scheme of neutrino mixing and
varies from three to six. 
The problem of neutrino masses and mixing is very important
for understanding of fundamental issues of elementary particle 
physics and astrophysics and has been treated in numerous review
articles, see e.g. \cite{val92,ver86,bil78,bil87,boe84,kay89,dol80}.

The investigation of  neutrino properties is a way to discover 
new physics beyond the SM. However, the problem of neutrino 
masses and mixing is still far of being solved. 
The finite mass of neutrinos is related to the problem of  lepton flavor 
violation. The SM strictly conserves lepton flavor but the GUT 
extensions of the
SM violate lepton flavor at some level. The lepton flavor violation 
has been discussed in several review articles 
\cite{ver86,her93,van93,dep95}. 

By using the flavor fields $\nu_L$, $(\nu_L)^c$, $\nu_R$ and $(\nu_R)^c$
(the indices L and R refer to the left-handed and right-handed chirality
states, respectively, and the superscript $c$ refers to the operation 
of charge conjugation) one can construct Dirac, Majorana and Dirac-Majorana 
neutrino mass Lagrangian's:
\begin{eqnarray}
{\cal L}^D & = & 
- \sum_{ll'} ~\overline{\nu_{l'R}}~ M^D_{l'l} ~\nu_{lL} + h.c., 
\label{mass.1} \\
{\cal L}^M & = & 
- \frac{1}{2} \sum_{ll'} ~\overline{(\nu_{l'L})^c} 
~M^M_{l'l}~ \nu_{lL} + h.c.,
\label{mass.2} \\
{\cal L}^{D+M} & = & 
-\sum_{ll'} ~[~ \frac{1}{2} ~\overline{(\nu_{l'L})^c} 
~(M^M_L)_{l'l} ~\nu_{lL} 
+ \frac{1}{2} ~\overline{\nu_{l'R}}~ (M^M_R)_{l'l}~ (\nu_{lR})^c 
+ \overline{\nu_{l'R}} ~(M^D)_{l'l}~ \nu_{lL}~ ] + h.c.
\label{mass.3}
\end{eqnarray}
Here the index $l$ and $l'$ take the values $e, \mu, \tau$.

The Dirac neutrino mass term ${\cal L}^D$ ($M^D$ is a complex
non-diagonal 3x3 matrix) couples ``active'' left-handed 
flavor fields $\nu_{lL}$ (weak interaction eigenstates) with the 
right-handed fields $\nu_{lR}$ which do not enter into the interaction
Lagrangian of the SM. 
In the case of the Dirac neutrino mass term
the three neutrino flavor field are connected with the 
three Dirac neutrino mass eigenstates as follows:
\begin{equation}
\nu_{lL} = \sum_{i=1}^3 ~U_{li} ~~\nu_{iL}.
\label{mass.4}
\end{equation}
Here U is a unitary matrix and $\nu_i$ is the field of the Dirac neutrino 
with mass $m_i$. The mixing of neutrinos generated by Dirac mass term 
is analogous to the mixing of quarks described by the 
Cabbibo-Kobayashi-Maskawa matrix.

It is evident that lepton flavor is violated in the theories
with a Dirac mass term, if the matrix $M^D$ is not diagonal,
i.e. additive partial lepton numbers of electronic ($L_e$), muonic
($L_\mu$) and tauonic ($L_\tau$) type are not conserved.  
However, the Dirac mass term is invariant in respect to the global
gauge transformation $\nu_L \rightarrow e^{i\Lambda} \nu_L$,
$\nu_R \rightarrow e^{i\Lambda} \nu_R$ ($\Lambda$ is arbitrary) what 
implies the conservation of the total lepton number $L=L_e+L_\mu+L_\tau$.
Processes like the $0\nu\beta\beta$-decay are forbidden in this scheme. 

The Majorana mass term ${\cal L}^M$ in Eq. (\ref{mass.2}) 
couples  neutrino states of given
chirality and its charge conjugate. It is obvious that the Majorana mass 
term ${\cal L}^M$ is not invariant under the global gauge transformation, 
i.e. does not conserve the total lepton number $L$. 
Consequently, the theories with ${\cal L}^M$ allow total lepton charge 
non-conserving processes like the $0\nu\beta\beta$-decay. 

The Dirac-Majorana mass term ${\cal L}^{D+M}$ in Eq. (\ref{mass.3}) is the 
most general neutrino mass term that does not conserve the total lepton
number $L$. $M^M_L$ and $M^M_R$ are complex non-diagonal symmetrical
3x3 matrices. The flavor neutrino fields are superpositions of six
Majorana fields $\nu_i$ with definite masses $m_i$ \cite{bil87},
\begin{eqnarray}
\nu_{lL} = \sum_{i=1}^6~ U_{l i} ~~\nu_{iL}, ~~~
(\nu_{lR})^c = \sum_{i=1}^6 ~U_{\overline{l} i}~~ \nu_{iL}. \label{mass.5}
\end{eqnarray}
The fields $\nu_i$ satisfy the Majorana condition 
\begin{equation}
\nu_i \xi_i = \nu_i^c = C ~{\overline{\nu}}_i^T,
\label{mass.6}
\end{equation}
where C denotes the charge conjugation and $\xi$ is a phase factor.

The advantage of the Dirac-Majorana mass mixing scheme is that it allows
to explain the smallness of the neutrino mass within the so called 
see-saw mechanism proposed by Yanagida, Gell-Mann, Ramond 
and Slansky \cite{yan79}. If the model involves the violation of 
lepton number at the large mass scale  like in GUT's,
the see-saw mechanism is the most natural way 
to obtain light neutrino masses. 
In the simplest case of 
one flavor generation the ${\cal L}^{D+M}$ takes the form
\begin{equation}
{\cal L}^{D+M}  = 
 - \frac{1}{2} ~
\left(
\matrix{ \overline{(\nu_{L})^c}~~ \overline{\nu_{R}}}
\right)~
\left(
\matrix{ m_L ~~ m_D \cr m_D ~~ m_R}
\right)~
\left(
\matrix{ \nu_{L} ~~ (\nu_{R})^c }
\right).
\label{mass.7}
\end{equation}
If we assume the left-handed Majorana mass $m_L$ is zero, the 
right-handed Majorana mass is of the GUT's scale 
($ M_{GUT} \approx 10^{12} GeV$ ) and the Dirac mass $m_D$ is of the 
order of the charged lepton and quark masses ($m_F << M_{GUT}$), 
we obtain after diagonalization one small and one large eigenvalues:
\begin{equation}
   m_1 ~\approx ~\frac{m^2_F}{M_{GUT}},  ~~~~~ m_2 ~ \approx ~ M_{GUT}. 
\label{mass.8}
\end{equation}
By considering the general case of three flavor generation we end up
with three light masses $m_i$ and three very heavy masses
$M_i$ ($m_i \approx (m^i_F)^2/M_i$). 

The information about the neutrino masses and neutrino mixing 
is obtained from different experiments.  The laboratory studies
of lepton charge conserving weak processes offer 
model independent neutrino mass limits that follow purely from kinematics. 
The most stringent bound 4.35 eV on the ``electron neutrino mass''  
has been obtained from the 
Troitsk tritium beta decay experiment \cite{bel95}
by investigation of the high energy part of the $\beta$ spectrum. 
The tritium beta decay experiments 
suffer from the anomaly given by the negative
average value of $m^2_\nu$. It means that instead of the expected
deficit of the events at the end of the spectrum some excess of events is 
observed. The PSI study of the pion decay gives the  strongest 
upper limit 170 KeV on the ``muon neutrino mass'' \cite{par96}. Further
improvement of this limit is restricted by the uncertainty in the
$\pi^-$ mass.  
By studying  the end point of the hadronic mass spectrum in the 
decay $\tau^- \rightarrow 2\pi^+3^-\nu_\tau$ an 
upper ``tau neutrino mass'' limit of 24 MeV was obtained \cite{par96}.

The experimental study  of neutrino
oscillations allows  to obtain interesting information about
some lepton flavor violating parameters. There are neutrino
mixing angles and differences of masses-squared $\Delta m^2$.   
In the neutrino oscillation experiments one searches for a deficit 
of some kind of neutrinos at some  distance from the source of neutrinos
$\nu_l$ (disappearance experiment) or one is looking for an appearance 
of neutrinos of a given kind $\nu_{l'}$ ($l'=e, \mu, \tau$) at some 
distance from the source of neutrinos of different kind $\nu_l$ ($l \ne l'$)
(appearance experiment). As a result of several oscillation experiments
limits have been set on relevant mixing and mass differences. 
Till now, only the LSND experiment \cite{ath95} found positive signals 
in favor of neutrino oscillations. The events can be 
explained by $\overline{\nu_\mu} \leftrightarrow \overline{\nu_e}$ 
oscillations with $\Delta m^2 \sim 1 eV^2$.

The second important indication in favor of neutrino masses 
and mixing comes from the Kamiokande \cite{fuk94}, IMB \cite{bec95} and 
Soudan \cite{goo96} experiments on the detection of atmospheric neutrinos.
The data of the Kamiokande collaboration can be explained by
${\nu_\mu} \leftrightarrow {\nu_e}$  or 
${\nu_\mu} \leftrightarrow {\nu_\tau}$ oscillations with  
$\Delta m^2 \sim 10^{-2} eV^2$.

The third most important indication comes from solar neutrino 
experiments (Homestake \cite{cle95}, Kamiokande \cite{hir91}, 
Gallex \cite{gal95} and SAGE \cite{abd94}). 
The neutrinos from the sun are detected by observation of the 
weak interaction induced reactions. The observed events still pose
a persisting puzzle being significantly smaller than the
values predicted by the Standard Solar Model SSM \cite{bah95}. 
These experimental data can be explained in the framework of MSW \cite{mih86}
matter effect for $\Delta m^2 \sim 10^{-5} eV^2$  or by vacuum
oscillations in the case of $\Delta m^2 \sim 10^{-10} eV^2$.

It is worthwhile to mention that
all the above mentioned existing indications in favor of neutrino mixing
cannot be described by any scheme with three massive neutrinos \cite{oka97}.
This fact implies that a
scheme of mixing with at least four massive neutrinos
(that include not only $\nu_e$, $\nu_\mu$, $\nu_\tau$ but at least
one sterile neutrino) has to be considered.

A prominent role among the neutrino mass experiments plays the 
$0\nu\beta\beta$-decay violating the total lepton number by two
units. A non-vanishing $0\nu\beta\beta$-decay rate requires neutrinos 
to be Majorana particles, irrespective which mechanism is used 
\cite{sche82}. This theorem has been generalized also for the case
of any realistic gauge model with the weak scale softly broken SUSY
\cite{hir97,hir97t}. 

The $0\nu\beta\beta$-decay is a second order process in the theory of weak
interaction and neutrino mixing enters into the matrix element 
of the process through the propagator
\begin{equation}
<{{\nu_{eL}(x_1)\nu}}^T_{eL,R}(x_2)>
 =  \sum_{k} U^L_{ek}  U^{L,R}_{ek} \xi_k  \frac{1- {\gamma}_5}{2}
<{{\nu_{k}(x_1)\overline{\nu}}}_k(x_2)>
\frac{1\mp{\gamma}_5}{2} C,
\label{mass.9}
\end{equation}
where the propagator of virtual neutrino $\nu_k$ 
($k = 1, 2 ...$) with Majorana mass
$m_k$ takes the form
\begin{equation}
<{{\nu_{k}(x_1)\overline{\nu}}}_k(x_2)> = 
\frac{1}{{(2\pi)}^4}
\int d^4 p \,
e^{-ip(x_1-x_2)}
\frac{\hat{p}+m_k}
{p^2-m^2_k}.
\label{mass.10}
\end{equation}
Eq. (\ref{mass.9}) can be derived as result of
Eq. (\ref{mass.6}) and the following relations:
\begin{eqnarray}
\nu_{eL} &= & \sum_k U^L_{ek} \nu_{kL} ~~~
(U^L_{ek}~ = ~U_{ek} ~ e^{-i\alpha_k}, e^{-2i\alpha_k}=\xi_k),
\label{mass.11}\\
\nu_{eR} &= & \sum_k U^R_{ek} \nu_{kR} ~~~
(U^R_{ek}~ = ~U^*_{\overline{e}k} ~ e^{i\alpha_k}~\xi_k).
\label{mass.12}
\end{eqnarray}
In the case the matrices $U^L$ and $U^R$ are not independent 
i.e. in the case of Dirac-Majorana  mass term, they obey 
the normalization and
orthogonality conditions \cite{bil87} 
\begin{eqnarray}
\sum^{2n}_{k=1} ~ U^L_{lk}~ (U^L_{l'k})^*  ~ = ~\delta_{ll'},
~~~\sum^{2n}_{k=1} ~ U^L_{lk}~ U^R_{l'k} ~\xi_k ~ = 0,
~~~\sum^{2n}_{k=1} ~ U^L_{lk}~ (U^R_{l'k})^*  ~ = ~\delta_{ll'}.
\label{mass.13}
\end{eqnarray}
From LEP data it follows from 
the measurement of the total decay width 
 of Z vector boson that the number of flavor neutrinos 
is equal to three \cite{par96}, i.e.
n=3 and $l,l' = e, \mu, \tau$.

If we suppose that the  light neutrino 
($m_k \ll $ few MeV) exchange is the dominant
mechanism for the $0\nu\beta\beta$-decay,  
the $0\nu\beta\beta$-decay amplitude is proportional to the 
following lepton number violating parameters:\\
i) If both current neutrinos are left handed one can separate 
an effective (weighted average) neutrino mass
\begin{equation}
<m> ~ = ~ \sum^{light}_k~ (U^L_{ek})^2 ~ \xi_k ~ m_k ~ = ~
\sum^{light}_k~ U_{ek}^2 ~ m_k.
\label{mass.14}
\end{equation}
We note that $<m>$ may be suppressed by a destructive interference
between the different contributions in the sum of Eq. (\ref{mass.14})
if CP is conserved.  In this case the mixing matrix satisfies 
the condition $U_{ek} = U^*_{ek}~\zeta_k$, where $\zeta_k = \pm i$ is the
CP parity of the Majorana neutrino $\nu_k$ \cite{bil84}. Then we have 
\begin{equation}
<m_\nu> ~ = ~\sum^{light}_k~ |U_{ek}|^2 ~ 
\zeta_{k} ~ m_k. 
\label{mass.15}
\end{equation}
ii) If both current neutrinos are of opposite chirality the amplitude is
proportional to the factor
\begin{equation}
\epsilon_{LR} ~ = ~ \sum^{light}_k~ U^L_{ek}~U^R_{ek} ~ \xi_k 
\label{mass.16}
\end{equation}
and the amplitude does not explicitly depend on the neutrino mass.
We note that in Eqs. (\ref{mass.14}) and (\ref{mass.16}) 
the summation is only over light neutrinos. 

If we consider only the heavy neutrino ($m_k \gg $ 1 GeV) exchange 
$0\nu\beta\beta$-decay mechanism and both current neutrinos to be 
left-handed, we can separate from the $0\nu\beta\beta$-decay amplitude 
the parameter
\begin{equation}
<m^{-1}_\nu> ~ = ~ \sum^{heavy}_k~ (U^L_{ek})^2 ~ \xi_k ~ \frac{1}{m_k}.
\label{mass.17}
\end{equation}
Here, the summation is only over heavy neutrinos. 

Most investigation study the $0\nu\beta\beta$-decay 
mechanism generated by the effective neutrino mass parameter given in 
Eq. (\ref{mass.14}). The value of this parameter can be 
determined in two ways.\\
i) One can extract $|<m_\nu>|$ from the best presently available 
experimental lower limits on the half-life of the $0\nu\beta\beta$-decay
after calculating the corresponding nuclear matrix elements.
>From the results of the $^{76}Ge$ experiment \cite{gue97} 
one finds $|<m_\nu>| <$ 1.1 eV \cite{simf97}. We note 
that a significant progress in search for $0\nu\beta\beta$-decay is
expected in the future. The Heidelberg-Moscow and NEMO collaborations
are planning to reach sensitivity of 0.1-0.3 eV for $|<m_\nu>|$.
\\
ii) One can use the constraints imposed 
by the results  of neutrino oscillation experiments
on $<m_\nu>$. Bilenky et al \cite{biln97} have shown that in a general scheme 
with three light Majorana neutrinos and a hierarchy of neutrino
masses (see-saw origin) 
the results of the first reactor long-baseline experiment 
CHOOZ \cite{apo97} and Bugey experiment \cite{ach95} imply: 
$|<m_\nu>| < 3\times 10^{-2} eV$. Thus the observation 
of the $0\nu\beta\beta$-decay with a half-life corresponding
to $|<m_\nu>| > 10^{-1} eV$ would be a signal of a non-hierarchical
neutrino mass spectrum or another mechanism for the violation
of lepton number, e.g. right-handed currents \cite{moh95,pan96},
$R_p \hspace{-1em}/\;\:$ SUSY \cite{moh86,ver87,hkk96,fae97,wo97} 
or others \cite{hir97,lpq96,pan97}.   

\section{Double beta decay}

In this Section the main formulas relevant for the 
$2\nu\beta\beta$-decay, Majorana neutrino exchange
and $R_p \hspace{-1em}/\;\:$  SUSY mechanisms of 
$0\nu\beta\beta$-decay are presented. We note that
the theory of the $2\nu\beta\beta$-decay and Majorana neutrino exchange
mechanism of $0\nu\beta\beta$-decay was discussed in details e.g.
in an excellent review of Doi et al \cite{doi85}.
The theory of the $R_p \hspace{-1em}/\;\:$  SUSY mechanism of
$0\nu\beta\beta$-decay was presented comprehensively in 
\cite{hkk96,fae97}. The readers are referred
to these publications for the details which are not covered in this
review.

\subsection{Two-neutrino mode}

The $2\nu\beta\beta$-decay is allowed in the SM, so the 
effects on the lepton number non-conservation can be neglected. 
In the most popular two nucleon mechanism of the $2\nu\beta\beta$ - decay 
process the beta decay Hamiltonian acquires the form:
\begin{equation}
{\cal H}^\beta(x)=\frac{G_{F}}{\sqrt{2}} 
\bar{e}(x)\gamma^\alpha (1-\gamma_5) \nu_{e}(x)~
j_{\alpha (x)} + {h.c.},
\label{two.1}
\end{equation}
where $e(x)$ and $\nu (x)$ 
are the field operators of the electron and neutrino, respectively. 
$j_{\alpha}(x)$ is free left-handed hadronic current.

The matrix element of this process takes the following form:
\begin{eqnarray}
<{f}|S^{{(2)}}|{i}> &=&\frac{(-{i})^{{2}}}{2}
{\left(\frac{G_{{F}}}{~\sqrt{2}}\right)}^{{2}}
L^{\mu\nu}(p_{{1}},p_{{2}},k_{{1}},k_{{2}}) 
J_{\mu\nu}(p_{{1}},p_{{2}},k_{{1}},k_{{2}}) 
\nonumber \\
&&-(p_{{1}}\leftrightarrow p_{{2}}) - 
(k_{{1}}\leftrightarrow k_{{2}}) + 
(p_{{1}}\leftrightarrow p_{{2}})
(k_{{1}}\leftrightarrow k_{{2}}),
\label{two.2} 
\end{eqnarray}
with
\begin{eqnarray}
J_{\mu\nu}(p_{{1}},p_{{2}},k_{{1}},k_{{2}})
=\int {{e}}^{{i}{{(p}}_{{1}}
{{+k}}_{{1}}{{)x}}_{{1}}}
{{e}}^{{i}{{(p}}_{{2}}
{{+k}}_{{2}}{{)x}}_{{2}}}\nonumber \\
{ _{{out}}<}p_{{f}}|T(J_\mu(x_{{1}})
J_\nu(x_{{2}}))|p_{{i}}>_{{in}} {d}x_{{1}} 
{d}x_{{2}}.
\label{two.3}
\end{eqnarray}
Here, $L^{\mu\nu}$ originates from the   lepton currents and 
$J_\mu(x)$ is the weak charged nuclear hadron current in the Heisenberg 
representation \cite{bil87,bil82}.   
$p_{{1}}$ and $p_{{2}}$ ($k_{{1}}$ and
$k_{{2}}$) are four-momenta of electrons (antineutrinos), 
$p_{{i}}$ and $p_{{f}}$ are four-momenta of the initial and final nucleus.

The matrix element in Eq.\ (\ref{two.2}) contains the contributions 
from two subsequent nuclear beta decay processes and 
$2\nu\beta\beta$-decay. They can be separated, if we write the
T-product of the two hadron currents as follows \cite{sim91}:
\begin{eqnarray}
T(J_\mu(x_{{1}}) J_\nu(x_{{2}})) 
=J_\mu(x_{{1}}) J_\nu(x_{{2}}) + 
\Theta(x_{{20}} - x_{{10}})
[J_\nu(x_{{2}}),J_\mu(x_{{1}})].
\label{two.4}
\end{eqnarray}
The product of two currents in the r.h.s. of Eq.\ (\ref{two.4}) 
is associated with two subsequent nuclear beta decay processes, which are 
energetically forbidden for most of the $2\nu\beta\beta$-decay isotopes. 
Thus the non-equal-time commutator of the two hadron currents
corresponds to $2\nu\beta\beta$-decay process. 

If standard approximations are assumed (only s-waves states of emitted 
electrons are considered, lepton energies are replaced with their
average value  ...) we have \cite{sim91} 
\begin{eqnarray}
J_{\mu\nu}(p_{{1}},
p_{{2}},k_{{1}},k_{{2}})=
-i2 M^{2\nu}_{{GT}}\delta_{\mu {k}}
\delta_{\nu {k}} ~~~~\nonumber \\
\times 2\pi\delta(E_{{f}}-E_{{i}}+
p_{{10}}+k_{{10}}+p_{{20}}+k_{{20}})
,~k=1,2,3,
\label{two.5}
\end{eqnarray}
where,
\begin{eqnarray}
M^{2\nu}_{GT} = <0^{{+}}_f|~
\frac{i}{2} \int_0^\infty 
[ A_{{k}}(t/2), A_{{k}}(-t/2) ] dt~ |0^{{+}}_i>, 
\label{two.6} 
\end{eqnarray}
with
\begin{eqnarray}
A_k (t) = e^{i H t} A_k (0) e^{- i H t} =
\sum^{\infty}_{{n=0}}\frac{(it)^{{n}}}{n!}
\overbrace{[H[H...[H}^{{n}~{times}}
,A_k(0)]...]].
\label{two.7} 
\end{eqnarray}
Here, $|0^{{+}}>_i$ and $|0^{{+}}>_f$ 
are the wave functions of the initial and  final  
nuclei with their corresponding energies 
$E_{{i}}$ and $E_{{f}}$, respectively. 
$\Delta$ denotes the average energy $\Delta = (E_{{i}}-E_{{f}})/2$. 
$A_{{k}}(0)$ is the Gamow-Teller transition operator $A_{{k}}(0)=\sum_{{i}} 
\tau^{{+}}_{{i}}(\vec{\sigma}_{{i}})_{{k}}$, k=1,2,3.

After performing the integration over the time, 
inserting a complete set  
of intermediate states $|1^+_n>$ with eigenenergies $E_{{n}}$
between the two axial currents in Eq.\ (\ref{two.6}) and
assuming that the nuclear states are eigenstates of the nuclear
Hamiltonian, one ends up  with the well-known form the Gamow-Teller 
transition matrix element in second order
\begin{eqnarray}
M^{2\nu}_{\text{GT}}&=&\sum_{\text{n}}\frac{
<0^{{+}}_{{f}}|A_{{k}}(0)|1^{{+}}_{{n}}>
<1^{{+}}_{{n}}|A_{{k}}(0)|0^{{+}}_{{i}}>}
{E_{{n}}-E_{{i}}+\Delta}.
\label{two.8}
\end{eqnarray}
So, $2\nu\beta\beta$-decay can be expressed in terms of 
single beta decay  transitions through virtual intermediate
states. The form of $M^{2\nu}_{GT}$ in Eq. (\ref{two.8}) is suitable for
approaches using states in the intermediate nucleus 
 (Intermediate Nucleus Approach = INA) to $2\nu\beta\beta$-decay process
like QRPA, RQRPA  and shell model methods, which construct the spectrum 
of the intermediate nucleus by  diagonalization. 

Under the introduced assumptions the inverse 
half-life for $2\nu\beta\beta$-decay 
can be written in the form with factorized lepton and nuclear parts:
\begin{equation}
[ T^{2\nu}_{1/2}(0^+ \rightarrow 0^+) ]^{-1} = 
G^{2\nu} ~ | M^{2\nu}_{GT}|^2,
\label{two.9}
\end{equation}
where $G^{2\nu}$ results from integration over the lepton phase space.
For the decays of interest $G^{2\nu}$ 
can be found e.g. in refs. \cite{hax84,doi85}.

It is worthwhile to notice that the form of $M^{2\nu}_{GT}$ in 
Eq. (\ref{two.6}) as a time integral 
is very useful for an analytical study of 
the different approximation schemes. It is obvious that the 
$2\nu\beta\beta$-decay operator should be at least
a two-body operator changing two neutrons into two protons. 
If the commutator of the two time 
dependent axial currents in Eq. (\ref{two.6})
is calculated without approximation for a nuclear Hamiltonian consisting 
of one- and two- body operators, one obtains an
infinite sum of many-body operators. However, if the single particle 
Hamiltonian approximation scheme is employed, the transition operator 
is zero \cite{simm97,sims97}. The QBA and renormalized QBA (RQBA) schemes 
imply for the $2\nu\beta\beta$-decay transition operator to be a constant.
It means that the $2\nu\beta\beta$-decay is a higher order process in the
boson expansion of the nuclear Hamiltonian and its higher order boson
approximations are important. This important phenomenon can not be 
seen within INA \cite{simm97,sims97}.
 
In view of the smallness of the nuclear matrix element $M^{2\nu}_{GT}$ in 
Eq. (\ref{two.8}) the electron $p_{1/2}$-wave Coulomb corrections
to the $2\nu\beta\beta$-decay amplitude for $0^+_{g.s} \rightarrow 0^+_{g.s.}$
transition have been examined.  It was done first 
within the closure approximation in \cite{sim88} with the conclusion
that the effect is small.
The more careful evaluation of the corresponding nuclear matrix
elements involves the summation over all virtual $0^-$ and $1^-$ states.
In the study performed by Krmpoti\'c et al \cite{bar95} it is argued 
this contribution to be significant. 
If the $p_{3/2}$-wave states
of outgoing electrons are considered, the nuclear transition 
can take place through virtual $2^-$ intermediate states.
However, it has been shown in \cite{civ96}
that such contribution to the $2\nu\beta\beta$-decay
half-life can be disregarded.

The inverse 
half-life of the $2\nu\beta\beta$-decay transition to the $0^+$ and $2^+$
excited states of the final nucleus is given as follows: 
\begin{equation}
[ T^{2\nu}_{1/2}(0^+ \rightarrow {\bf J}^+) ]^{-1} = 
G^{2\nu}(J^+) ~ | M^{2\nu}_{GT}(J^+)|^2,
\label{two.10}
\end{equation}
with 
\begin{eqnarray}
M^{2\nu}_{\text{GT}}(J^+) = \sum_{\text{n}}\frac{
<0^{{+}}_{{f}}|A_{{k}}(0)|1^{{+}}_{{n}}>
<1^{{+}}_{{n}}|A_{{k}}(0)|0^{{+}}_{{i}}>}
{[{E_{{n}}-E_{{i}}+\Delta}]^s},
\label{two.11}
\end{eqnarray}
where $s = 1$ for $J = 0$ and $ s = 3 $ for $J = 2$.
These transitions are disfavored by smaller phase space factors  
 $G^{2\nu}(J^+) $ 
due to a smaller available energy release for these processes
\cite{doi83}.

\subsection{Neutrinoless mode}

The presently favored left-right symmetric models 
of Grand Unifications pioneered 
by Mohapatra, Pati and Senjanovi\'c \cite{moh75} are 
based on a very interesting  idea. It is assumed that parity
violation is a low energy phenomenon and that parity conservation is
restored above a certain energy scale. Parity could be violated
spontaneously in the framework of gauge theories built on the
electroweak $SU(2)_R \otimes SU(2)_L \otimes U(1)$ gauge group
\cite{moh81}. In addition the left-right group structure can
be derived from GUT groups SO(10) \cite{fri81} and the small neutrino
masses can be explained with help of the see-saw mechanism in a
natural way. 

The left-right models contain the known vector bosons $W^\pm_L$ 
(81 GeV) mediating the left-handed weak interaction and 
hypothetical  vector bosons $W^\pm_R$ responsible for a right-handed
weak interaction:
\begin{eqnarray} 
W^{\pm}_1  =  \cos\zeta W^{\pm}_L + \sin\zeta W^{\pm}_R, ~ ~~~
W^{\pm}_2  =  -\sin\zeta W^{\pm}_L + \cos\zeta W^{\pm}_R.
\label{neu.1}
\end{eqnarray}
The ``left- and right-handed'' vector bosons are mixed if the mass eigenstates
are not identical with the weak eigenstates. 
Since we have not seen a right-handed weak interaction, the mass of a heavy
vector boson must be considerably larger than the mass of the light vector
boson, which is responsible mainly for the left-handed weak interaction. 

Within the left-right models 
the weak beta decay Hamiltonian constructed by Doi,
Kotani, Nishiura and Takasugi \cite{doi83} takes the form
\begin{equation}
{\cal H}^\beta = \frac{G_F}{\sqrt{2}}
[ j_{L\mu}J^{\mu +}_L + \kappa j_{L\mu}J^{\mu +}_R + 
\eta j_{R\mu}J^{\mu +}_L + \lambda j_{R\mu}J^{\mu +}_R ] + h.c.   
\label{neu.2}
\end{equation} 
Here, $\kappa = \eta = \tan \zeta$ and $\lambda = (M_1/M_2)^2$
are dimensionless coupling constants for different parts of the  
right handed weak interaction.
$M_1$ and $M_2$ are respectively the masses of the vector bosons
$W_1$ and $W_2$. 
The left- and right- handed leptonic currents are
\begin{equation}
j^\mu_L(x) = \overline{e}(x) \gamma^\mu (1-\gamma_5) \nu_{eL}(x),
~~~
j^\mu_R(x) = \overline{e}(x) \gamma^\mu (1+\gamma_5) \nu_{eR}(x).
\label{neu.3}
\end{equation}
The left- and right- handed current neutrinos ($\nu_{eL}$ and $\nu_{eR}$)
are given in Eqs. (\ref{mass.11}) and (\ref{mass.12}). The nuclear
currents are assumed to be 
\begin{eqnarray}
(J^{0 +}_{L,R}(\mathbf{x}), {\mathbf{J}}^{0 +}_{L,R}(\mathbf{x}))
= \sum_{n=1}^A \tau^+_n \delta ( \mathbf{x}-{\mathbf{r}}_n ) 
(g_V\mp g_A C_n, \mp g_A {\mathbf{\sigma}}_n + g_V {\mathbf{D}}_n ),
\label{neu.4}
\end{eqnarray}
where $C_n$ and ${\mathbf{D}}_n$ are nucleon recoil terms \cite{doi85}. 
The $\eta$ term arises when the chiralities of the quark hadronic 
current match those of the leptonic current, i.e. there are of 
left-right combination (this is possible due W-boson mixing). 
The $\lambda$ term arises when both the  quark hadronic 
current and leptonic current are right-handed [ see Fig. \ref{figzero.1}].

Let suppose that $0\nu\beta\beta$-decay is generated by the weak
interaction Hamiltonian in Eq. (\ref{neu.2}) and that only exchange of light
neutrinos is considered. Then, the matrix element for this process depends on 
the three effective lepton number violating parameters 
$<m_\nu>$ (defined in Eq. (\ref{mass.14})), 
$< \lambda >$ and $<\eta >$:
\begin{equation}
<\lambda > = \lambda ~~\epsilon_{LR} = \lambda ~
\sum^{light}_k~ U^L_{ek}~U^R_{ek} ~ \xi_k, ~~~  
<\eta >  = \eta ~\epsilon_{LR} = \eta ~
\sum^{light}_k~ U^L_{ek}~U^R_{ek} ~ \xi_k.
\label{neu.5}
\end{equation} 
We note that the $0\nu\beta\beta$-decay amplitude converges to zero
in the limit of zero neutrino mass. It is apparent by $<m_\nu> = 0$
in this case.  If all neutrinos are massless or are massive 
but light , $<\lambda>$ and $<\eta>$ vanish  due 
to the orthogonality condition in Eq. (\ref{mass.13}).  

The inverse half-life  for $0\nu\beta\beta$-decay 
expressed in terms of the lepton number violating parameters 
is given by \cite{doi85}
\begin{eqnarray}
[T^{0\nu}_{1/2}]^{-1} & = & |M^{0\nu}_{GT}|^2 
[ C_{mm} (\frac{m_\nu }{m_e})^2 + 
C_{\lambda \lambda} <\lambda >^2 + C_{\eta\eta} <\eta >^2  +\nonumber \\ 
&&C_{m \lambda} |<\lambda>| \frac{|<m_\nu >|}{m_e} \cos\Psi_1 + 
C_{m \eta} |<\eta>|  \frac{|<m_\nu >|}{m_e} \cos\Psi_2 + \nonumber \\
&&C_{\lambda\eta} |<\lambda>| |<\eta>| \cos(\Psi_1-\Psi_2) ],
\label{neu.6}
\end{eqnarray}
where $\Psi_1$ and $\Psi_2$ are the relative phases between 
$<m_\nu>$ and $<\lambda >$  and $<\eta>$, respectively.
The coefficients $C_{xy}$ are expressed as combination of 8 nuclear 
matrix elements and 9 kinematical factors:
\begin{eqnarray}
C_{mm}&=&(1-\chi_F)~ G_{01}, ~~~~
C_{m \eta} = (1-\chi_F )[ \chi_{2+} G_{03} - \chi_{1-} G_{04} -
{\chi}_{P} G_{05} + \chi_{R} G_{06} ], \nonumber \\
C_{m \lambda}&=& -(1-\chi_F )[\chi_{2-} G_{03} - \chi_{1+} G_{04} ], ~~
C_{\lambda \lambda} = [ \chi^2_{2-} G_{02} + 
\frac{1}{9} \chi^{2}_{1-} G_{04} - \frac{2}{9} \chi_{1+}\chi_{2-} G_{03} ] 
\nonumber \\
C_{\eta \eta}&=&[ \chi^2_{2+} G_{02} + \frac{1}{9}\chi^2_{1-} G_{04} +
- \frac{2}{9}\chi_{1-}\chi_{2+} G_{03} + \chi^2_P G_{08} - 
\chi_P \chi_R G_{07} + \chi^2_R G_{09} ], \nonumber \\
C_{\lambda \eta}&=&-2[ \chi_{2-}\chi_{2+} G_{02} - 
\frac{1}{9}(\chi_{1+}\chi_{2+} + \chi_{2-}\chi_{1-}) G_{03} + 
\frac{1}{9} \chi_{1+}\chi_{1-} G_{04} ],
\label{neu.7}
\end{eqnarray} 
with
\begin{equation}
\chi_{1\pm} = \pm 3 {{\chi}}_{{F'}} + {{\chi}}_{{GT'}} -
6 {\chi_{{T'}}}, ~~~
\chi_{2\pm} = \pm  {{\chi}}_{F\omega} + {{\chi}}_{GT\omega} -
\frac{1}{9} \chi_{1\pm}.
\label{neu.8}
\end{equation}
The numerical values of the kinematical factors $G_{0i}$ ($i=1,9$)
can be found e.g. in \cite{doi85,pan96}. The nuclear matrix elements 
within closure approximation are of the form
\begin{equation}
X_I =
\langle 0^+_f|\sum_{i\neq j} \tau_i^+ \tau_j^+ 
{\cal O}_{I}({\mathbf{r}}_i,{\mathbf{r}}_j,
{\mathbf{\sigma}}_i,{\mathbf{\sigma}}_j) | 0^+_i \rangle 
\label{neu.9}
\end{equation}
with $X_I = M^{0\nu}_I$ ($I = GT$) or  $X_I = \chi_I$ 
($I = GT$, $F$, $GT\omega $, $F\omega $, 
${GT'} $, ${F'}$, ${T'}$, $P$, $R$). 
The radial and spin-angular
dependence of the transitions operators ${\cal O}_I $ 
is given in Table \ref{tabneu.1}.
We mention that the inaccuracy coming from the closure approximation 
in completing the sum over virtual nuclear states is small 
(about $15 \%$)  \cite{mut95},
since the average neutrino momentum  is large compared to the
average nuclear excitation energy 
\cite{tom91}. The nuclear matrix elements derived without
closure approximation can be found in \cite{ver90,pan92,krm94,bar97}. 
The expressions for the $0\nu\beta\beta$-decay matrix elements  obtained
within the relativistic quark confinement model are given in 
\cite{suh91}.

We note that a more general expression for the inverse half-life including 
also the
heavy neutrino exchange mechanism can be found in \cite{pan92}. We shall
discuss this type of $0\nu\beta\beta$-decay mechanism 
in context of the $R_p \hspace{-1em}/\;\:$-SUSY mechanisms.

\subsection{Majoron-Neutrinoless mode} 

In many extensions of  the standard model 
\cite{chi81,gon89,rom92,ber92} 
appears a physical Goldstone boson called Majoron, which is a 
light or massless boson with a very tiny coupling to neutrinos. 
The Majoron offers a new possibility
for looking for a signal of new physics in double beta decay 
experiments. 

The Majoron emitting mode of double beta decay 
($0\nu\beta\beta\phi$-decay) [see Eq. (\ref{int.3})] has been proposed
by Georgi, Glashow and Nussinov \cite{geo81} 
within the model introduced by Gelmini and Roncadelli \cite{gel81}. 

Today, we know several Majoron models based on different motivations
leading to a different form of the 
Majoron-neutrino interaction Lagrangian. The classical  
Majoron Models \cite{chi81,gel81,aul82} 
suggest that the interaction of the so called ordinary Majoron 
with neutrinos is of the form: 
\begin{equation}
{\cal L}_{\phi\nu\nu} = -\frac{1}{2}\sum_{ij} 
{\overline{\nu}}_{i} (a_{ij}P_L +b_{ij} P_R) \nu_j \phi^* + h.c.,
\label{maj.1}
\end{equation}
where $P_{R,L}= 1/2(1\pm\gamma_5)$.
If the usual approximation $m_{i,j} \ll q \approx p_F
\approx {\cal O}(100 MeV)$ valid for light neutrinos is assumed,
one can write for the inverse half-life of the
$0\nu\beta\beta\phi$-decay  
\begin{equation}
[T^{\phi 0\nu}_{1/2}]^{-1} = |<g>|^2  |M^{0\nu}_{GT}|^2 |1-\chi_F|^2 G_B, 
\label{maj.2}
\end{equation}
with the effective Majoron coupling constant
\begin{equation}
<g> = \sum_{i j}^{light} U^R_{ei} U^R_{ej} b_{ij}. 
\label{maj.3}
\end{equation}
We see that $0\nu\beta\beta\phi$-decay is ruled by the same nuclear
matrix elements as the light Majorana neutrino mass mechanism 
of the $0\nu\beta\beta$-decay. The numerical value of the kinematical
factors $G_B$ for different nuclei can be found e.g. in \cite{doi85}.

Recently, a new class of 
Majoron models \cite{bur93,bam95,car93} have been considered, 
which introduce a new types of Majorons carrying 
leptonic charge. The theoretical analysis of the charged Majoron modes
of double beta decay have been performed in \cite{hir96,bark95}.
It has been found that within the new Majoron models 
unobservable  small decay rates for the Majoron-emitting
$0\nu\beta\beta\phi$-decay are expected.

\subsection{SUSY neutrinoless mode}

Supersymmetry has been introduced to solve the so called 
``hierarchy problem'' of the unified theories \cite{chan88}. 
It is the only known symmetry which can stabilize the elementary Higgs boson
mass with respect to otherwise uncontrollable radiative corrections. 
Unlike the symmetries of ordinary particle physics, that  relate 
the particles of the same spin, the SUSY is relating bosons to fermions 
and vice-versa and requires the existence of new supersymmetric particles.
To each ordinary particle like leptons (neutrino, electron), quarks, 
gauge bosons (W, Z, photon), Higgs bosons  corresponds  
a  superpartner with different spin
(bosons replacing fermions and vice versa), 
i.e.  slepton (sneutrino, selectron), 
squark, gaugino (wino, zino, photino), higgsino. SUSY
is not a good  symmetry in the nature. Supersymmetry has to be 
broken otherwise the superpartners would possess the same masses as their 
ordinary partners. This would change drastically the phenomenology of
particle physics.

The simplest SUSY extension of the SM is the minimal supersymmetric 
standard model (MSSM). The Lagrangian of the MSSM conserves a new quantity
called R-parity:
\begin{equation}
R = (-1)^{(3B + L + 2S)}.
\end{equation}
Here, B, L and S are respectively the baryon number, lepton number and 
spin. The R-parity has been imposed guarantee the baryon and lepton number
conservation. Ordinary particles have R-parity even and their supersymmetric 
partners R-parity odd. However, R-parity conservation  is not required by 
gauge invariance or supersymmetry and might be broken explicitly or 
spontaneously at the Planck scale ($R_p \hspace{-1em}/\;\:$ SUSY) 
\cite{rom92}. 

The $R_p \hspace{-1em}/\;\:$ SUSY
is an other way of the lepton number violation in addition to the  Majorana 
neutrino mass term. It has interesting phenomenological consequences. 
The lepton flavor violation in the context of SUSY models has been
discussed extensively in the literature \cite{val91,rom91,leo86,kos89}.
The collider experiments \cite{roy92}, low-energy processes \cite{bar89}, 
matter stability \cite{zwi83,wei82,bar86} and cosmology 
\cite{cam91,moh87,dre91} offer important constraints on   
$R_p \hspace{-1em}/\;\:$ SUSY theories.

The $0\nu\beta\beta$-decay induced by exchange of superparticles has been 
studied within the $R_p \hspace{-1em}/\;\:$ MSSM \cite{hkk96,fae97,wo97}.
The relevant $R_p \hspace{-1em}/\;\:$ violating part of the superpotential
takes the form:
\begin{equation}
W_{R_p \hspace{-0.8em}/\;\:}~ = ~\lambda'_{ijk}~ L_i ~ Q_j ~{\bar D}_k.
\label{susy.1}
\end{equation}
Here $L$ and  $Q$ are respectively lepton and quark
doublet superfields while  ${\bar D}$  denotes ${\em down}$ quark 
singlet superfields. $\lambda'_{ijk}$ is a coupling constant and 
indices $i,j, k$ denote generations. 

The nuclear ${0\nu\beta\beta}$-decay
is triggered by the ${0\nu\beta\beta}$ quark transition
$d + d\rightarrow u + u + 2 e^-$. If this lepton number 
violating process ($\Delta L_e = 2$) is induced by heavy SUSY particles
and heavy neutrinos in a virtual intermediate state, one can write for 
the corresponding effective quark-electron Lagrangian 
\cite{fae98}:
\begin{eqnarray}
 {\cal L}^{\Delta L_e =2}_{eff}\ =
\frac{G_F^2}{2 m_{_p}}~ \bar e (1 + \gamma_5) e^{\bf c}~ 
\left[\eta_{PS}~ J_{PS}J_{PS} 
- \frac{1}{4} \eta_T ~   J_T^{\mu\nu} J_{T \mu\nu} 
+ \eta_N ~ J^\mu_{VA} J_{VA \mu} \right].
\label{susy.2}
\end{eqnarray}
The color singlet hadronic currents are 
$J_{PS} =   {\bar u}^{\alpha} (1+\gamma_5) d_{\alpha}$, 
$J_T^{\mu \nu} = {\bar u}^{\alpha} 
\sigma^{\mu \nu} (1 + \gamma_5) d_{\alpha}$, 
$J^\mu_{AV} = {\bar u}^{\alpha} \gamma^\mu (1-\gamma_5) d_{\alpha}$, 
where $\alpha$ is a color index and 
$\sigma^{\mu \nu} = (i/2)[\gamma^\mu , \gamma^\nu ]$.
Compared to Ref. \cite{hkk96} the 
Lagrangian in Eq. (\ref{susy.2}) contains properly taken into account
contribution of the color octet currents.

The effective lepton-number violating parameters $\eta_{PS}$
and $\eta_{T}$  in Eq.\ (\ref{susy.2}) accumulate fundamental parameters of 
$R_p \hspace{-1em}/\;\:$ MSSM as follows:
\begin{eqnarray}
\label{etaq}
\eta_{PS} &=&  \eta_{\chi\tilde e} + \eta_{\chi\tilde f} +
\eta_{\chi} + \eta_{\tilde g} + 7 \eta_{\tilde g}^{\prime}, \\
\label{eta}
\eta_{T} &=& \eta_{\chi} - \eta_{\chi\tilde f} + \eta_{\tilde g}
- \eta_{\tilde g}^{\prime},
\end{eqnarray}
with
\begin{eqnarray}
\label{eta_g}
\eta_{\tilde g} &=& \frac{\pi \alpha_s}{6}
\frac{\lambda^{'2}_{111}}{G_F^2 m_{\tilde d_R}^4} \frac{m_P}{m_{\tilde g}}\left[
1 + \left(\frac{m_{\tilde d_R}}{m_{\tilde u_L}}\right)^4\right]\\
\eta_{\chi} &=& \frac{ \pi \alpha_2}{2}
\frac{\lambda^{'2}_{111}}{G_F^2 m_{\tilde d_R}^4}
\sum_{i=1}^{4}\frac{m_P}{m_{\chi_i}}
\left[
\epsilon_{R i}^2(d) + \epsilon_{L i}^2(u)
\left(\frac{m_{\tilde d_R}}{m_{\tilde u_L}}\right)^4\right]\\
\eta_{\chi \tilde e} &=& 2 \pi \alpha_2
\frac{\lambda^{'2}_{111}}{G_F^2 m_{\tilde d_R}^4}
\left(\frac{m_{\tilde d_R}}{m_{\tilde e_L}}\right)^4
\sum_{i=1}^{4}\epsilon_{L i}^2(e)\frac{m_P}{m_{\chi_i}},\\
\eta'_{\tilde g} &=& \frac{\pi \alpha_s}{12}
\frac{\lambda^{'2}_{111}}{G_F^2 m_{\tilde d_R}^4}
\frac{m_P}{m_{\tilde g}}
\left(\frac{m_{\tilde d_R}}{m_{\tilde u_L}}\right)^2,\\
\label{eta_end}
\eta_{\chi \tilde f} &=& \frac{\pi \alpha_2 }{2}
\frac{\lambda^{'2}_{111}}{G_F^2 m_{\tilde d_R}^4}
\left(\frac{m_{\tilde d_R}}{m_{\tilde e_L}}\right)^2
\sum_{i=1}^{4}\frac{m_P}{m_{\chi_i}}
\left[\epsilon_{R i}(d) \epsilon_{L i}(e)  + \right.\\ \nonumber
&+& \left.\epsilon_{L i}(u) \epsilon_{R i}(d)
\left(\frac{m_{\tilde e_L}}{m_{\tilde u_L}}\right)^2
+ \epsilon_{L i}(u) \epsilon_{L i}(e)
\left(\frac{m_{\tilde d_R}}{m_{\tilde u_L}}\right)^2
\right].
\end{eqnarray}
where  
$\alpha_2 = g_{2}^{2}/(4\pi)$ and $\alpha_s = g_{3}^{2}/(4\pi)$ are
$SU(2)_L$ and $SU(3)_c$ gauge coupling constants.
$m_{\tilde u_L}$, $m_{\tilde d_R}$,
$m_{\tilde g}$ and $m_{\chi}$ are masses
of the u-squark, d-squark, 
gluino $\tilde g$ and of the lightest neutralino $\chi$,
respectively. The neutralino is  linear superposition 
of the gaugino and higgsino fields:
$\chi = \alpha_{\chi} \tilde{B} +  \beta_{\chi} \tilde{W}^{3} +
\delta_{\chi} \tilde{H}_{1}^{0} + \gamma_{\chi} \tilde{H}_{2}^{0}$.
Here $\tilde{W}^{3}$ and $\tilde{B}$ are neutral
$SU(2)_L$ and $U(1)$ gauginos while $\tilde{H}_{2}^{0}$,
$\tilde{H}_{1}^{0}$
are higgsinos which are the superpartners of the two neutral Higgs boson
fields $H_1^0$ and $H_2^0$
with a weak hypercharge $Y=-1, \ +1$, respectively.
The mixing coefficients $\alpha_{\chi},\beta_{\chi},\gamma_{\chi},
\delta_{\chi}$ can be obtained from diagonalization of the $4\times 4$
neutralino mass matrix \cite{Haber}.
Neutralino couplings are defined as \cite{Haber}
$\epsilon_{L\psi} = - T_3(\psi) \beta_{\chi} +
\tan \theta_W \left(T_3(\psi) -  Q(\psi)\right) \alpha_{\chi}$,
$\epsilon_{R\psi} = Q(\psi) \tan \theta_W \alpha_{\chi}$.
Here $Q $ and $ \ T_3$ are the electric charge and the weak
isospin of the fields $\psi = u, d, e$.
The relevant Feynman diagrams associated with 
gluino and neutralino contributions 
to the $0\nu\beta\beta$-decay are drawn in Fig. \ref{figsus.1}.

The effective lepton-number violating parameters $\eta_{N}$
in Eq.\ (\ref{susy.2}) has origin in the exchange of heavy Majorana 
neutrino  and takes the form: 
$\eta_N = m_p <m^{-1}_\nu >$. Here,  $m_p$ is the mass of proton and 
$<m^{-1}_\nu >$ is 
the inverse value of the effective heavy Majorana neutrino mass 
defined in Eq. (\ref{mass.17}).

For the calculation of the $0\nu\beta\beta$-decay nuclear transitions 
the quark-lepton interaction in Eq. (\ref{susy.2}) has to be
reformulated in terms of effective hadron-lepton interaction. 
The quarks undergoing $R_p \hspace{-1em}/\;\:$ SUSY transition
can be incorporated in two nucleons (the well-known two-nucleon mode),
one intermediate pion and one nucleon  (so called one-pion mode) or 
in two intermediate pions (so called two-pion mode).  These 
three possibilities are drawn in Fig. \ref{figsus.2}. We note that 
intermediate SUSY partners are heavy particles and thus 
in the two-nucleon mode the two-decaying neutrons must come very close
to each other. This contribution is therefore
suppressed by the short range nucleon-nucleon repulsion. Therefore, it is
expected that the pion-exchange mechanism dominates over the
conventional two-nucleon one. The scale on which the pion contribution
is enhanced compared to the two-nucleon mode is the ratio of the
nucleon form factor cut-off  parameter compared to
the pion mass. In addition, the relative 
importance of the two-nucleon mode, of the one-pion mode and of 
the two-pion mode
depends on the evaluation of the hadronic matrix element 
$<\pi^+ | J_i J_i |\pi^- >$ ($i = P,S,T$), which is model dependent.
The values of the corresponding structure coefficients calculated 
within the vacuum insertion approximation (VIA) and within the  
non-relativistic quark model (QM) differs from 
each other by a factor of three \cite{fae97}. It is worthwhile to notice
that the importance of the pion-exchange currents in the 
$0\nu\beta\beta$-decay was first been pointed out by Pontecorvo
\cite{pon68}. 

The half-life for the neutrinoless double beta decay regarding all the three
possibilities of hadronization of the quarks can be written in the
form \cite{fae98}
\begin{eqnarray}
\big[ T_{1/2}^{0\nu}(0^+ \rightarrow 0^+) \big]^{-1}~~ =
~~~~~~~~~~~~~~~~~~~~~~~~~~~~~~~~~~~~~~~~~~~~~~~~\nonumber \\
G_{01} \left | \eta_{T} {\cal M}_{\tilde q}^{2N}
 +  (\eta_{PS}-\eta_{T}) {\cal M}_{\tilde f}^{2N} + \frac{3}{8}
(\eta_{T} + \frac{5}{8} \eta_{PS}) {\cal M}^{\pi N} 
+\eta_N {\cal M}_N \right |^2\,
\label{susy.3}
\end{eqnarray}
where 
\begin{eqnarray}
{\cal M}^{2N}_{\tilde q} &=&  c_A \Big[
\alpha^{(0)}_{V-\tilde{q}} {\cal M}_{F N} + 
\alpha^{(0)}_{A-\tilde{q}} {\cal M}_{GT N} +
\alpha^{(1)}_{V-\tilde{q}} {\cal M}_{F'} + 
\alpha^{(1)}_{A-\tilde{q}} {\cal M}_{GT'} +
\alpha_{T-\tilde{q}} {\cal M}_{T'} \Big]\,, \nonumber \\
{\cal M}^{2N}_{\tilde f} &=& c_A \Big[
\alpha^{(0)}_{V-\tilde{f}} {\cal M}_{F N} + 
\alpha^{(0)}_{A-\tilde{f}} {\cal M}_{GT N} +
\alpha^{(1)}_{V-\tilde{f}} {\cal M}_{F'} + 
\alpha^{(1)}_{A-\tilde{f}} {\cal M}_{GT'} +
\alpha_{T-\tilde{f}} {\cal M}_{T'} \Big]\,, \nonumber \\
{\cal M}^{\pi N} &=& c_A \Big[
 \frac{4}{3}\alpha^{1\pi}\left(M_{GT-1\pi} + M_{T-1\pi} \right)
      +
      \alpha^{2\pi}\left(M_{GT-2\pi} + M_{T-2\pi} \right)\Big]\,,
\nonumber \\
{\cal M}_N &=&  c_A \Big[ 
\frac{g^2_V}{g^2_A} {\cal M}_{F N} - {\cal M}_{GT N} \Big],
~~~~c_A = \frac{m_{_p}}{ m_e}\left(\frac{m_A}{m_{_p}}\right)^2.
\label{susy.5}
\end{eqnarray}
Here $G_{01}$ is the standard phase space factor 
introduced in Eq. (\ref{neu.7})
and $m_A = 850$ MeV is the nucleon form factor cut-off (for all nucleon
form factors the dipole shape with the same cut-off is considered).
The nucleon structure coefficients $\alpha 's$ entering 
nuclear matrix elements of the two-nucleon mode are given in 
Table \ref{tabsus.1}. 
The structure coefficient of the one-pion  $\alpha^{1\pi}$
and two-pion mode $\alpha^{2\pi}$ are \cite{fae97,fae98}:
$\alpha^{1\pi} = -0.044$ and $\alpha^{2\pi} = 0.2$ (VIA), 0.64 (QM). 
The partial nuclear matrix elements of the
$R_p \hspace{-1em}/\;\:$ SUSY mechanism for the $0\nu\beta\beta$-decay 
appearing in Eq. (\ref{susy.5})
are of the following form:
\begin{equation}
{\cal M}_I =
\langle 0^+_f|\sum_{i\neq j} \tau_i^+ \tau_j^+ 
{\cal O}_{I}({\mathbf{r}}_i,{\mathbf{r}}_j,
{\mathbf{\sigma}}_i,{\mathbf{\sigma}}_j) | 0^+_i \rangle , 
\label{susy.8}
\end{equation}
where $I =$ $F N$, $GT N$, $F'$, $GT'$, $T'$, $GT-i\pi$ and 
$T-i\pi$ (i=1,2).  The corresponding operators ${\cal O}_{I}$ are
given in Table \ref{tabneu.1}.

After the nuclear matrix elements are evaluated one can extract 
the constraints on the combination of lepton number violating parameters
$\eta_{PS}$, $\eta_{T}$ and 
$\eta_{N}$ from the
non-observation of the $0\nu\beta\beta$-decay [see Eq. (\ref{susy.3})]. 
It is worthwhile to notice that it is a common practice to
consider separatelly $R_p \hspace{-1em}/\;\:$ SUSY and
heavy neutrino exchange mechanisms for the $0\nu\beta\beta$-decay. 
Then, the experimental lower bound $T^{exp}_{1/2}$ for the half-life
provides the following constraint on  $\eta_N$ and
$\eta_{_{SUSY}}$ parameters:
\begin{equation}
\eta_{N} \leq \eta^{exp}_{N} = \frac{1}{|{\cal M}_N |} 
\frac{1}{\sqrt{G_{01} T^{exp}_{1/2}}},
\end{equation}
\begin{eqnarray}
\label{con}
\eta_{_{SUSY}} \equiv  \frac{3}{8}(\eta^T + \frac{5}{3} \eta^{PS})
 \leq \eta^{exp}_{_{SUSY}} = \frac{1}{|{\cal M}^{\pi N}|} 
\frac{1}{\sqrt{G_{01} T^{exp}_{1/2}}}.
\label{limit.susy}
\end{eqnarray}
where we assumed the dominance of the two-pion mode 
(, which seems to be very probable). Then we have for the 
 1st generation Yukawa coupling constant ${\acute{\lambda}}_{111}$ 
under the phenomenologically viable simplifying assumptions 
(the lightest neutralino is B-ino dominant, 
$m_{\tilde q} \geq m_{\tilde e}/2$)  
\cite{fae98}:
\begin{eqnarray}
\lambda'_{111} \le 1.8
\sqrt{{\eta}^{exp}_{_{SUSY}}}
\Big({m_{\tilde q}\over{100 ~{\text{GeV}}}} \Big)^2
 \Big({m_{\tilde g}\over{100 ~{\text{GeV}}}} \Big)^{1/2},
\label{susy.9}\\
\lambda'_{111} \le 12.5
\sqrt{{\eta}^{exp}_{_{SUSY}}}
\Big({m_{\tilde e}\over{100 ~{\text{GeV}}}} \Big)^2
 \Big({m_{\chi}\over{100 ~{\text{GeV}}}} \Big)^{1/2}.
\label{susy.10}
\end{eqnarray}
We note that the running QCD coupling constant $\alpha_s(Q)$
was taken at the scale $Q = 1$GeV with the normalization on the
world average value $\alpha_s(M_Z) = 0.120$ \cite{par96}.

There is also a possibility to get useful constraints on other
parameters of the SUSY models by (e.g. B-L violating sneutrino mass)  
considering the R-parity conserving SUSY mechanism for 
$0\nu\beta\beta$-decay.  A class of such mechanisms have been 
considered in \cite{hir97}. However, the constraint on the sneutrino mass
spectrum
from the double beta decay experiment has been found less stringent then the 
constraint from the best laboratory limits on the neutrino masses.

\section{Nuclear models}

\subsection{Shell model}

The shell model is giving a very satisfactory description of the 
observed properties of the low-lying states of nuclei. The microscopic 
shell model wave functions posses a set of quantum numbers
reflecting the symmetries of the strong interaction. This 
is an advantage in respect to the QRPA
(Quasiparticle Random Phase Approximation), where the proton 
and neutron numbers are only conserved in average. 
However,  for the medium and heavy nuclei the
shell model space increases so drastically, that reliable calculations
are impossible beyond the pf-shell. But even relative large spaces 
in pf-shell nuclei can not reliably describe the states in the giant
Gamow-Teller resonance region, where most of the $\beta^-$ strength 
is concentrated. Thus even in lighter nuclei 
the shell model studies of the $2\nu\beta\beta$-decay
transitions are based on the assumption that the contribution of 
higher lying $\beta^-$ strengths to $2\nu\beta\beta$-decay matrix element
is negligible, e.g. due to the small overlap with the $\beta^+$ strength 
distribution and due to the large energy denominator. 

Calculation of the $0\nu\beta\beta$-decay matrix elements 
requires only the knowledge  of the initial and final nuclear
wave functions as the exchange potential depends weakly on 
details of
nuclear structure. It is worthwhile to notice that
it is almost impossible
 to construct even the wave functions of the initial 
and final nuclei for nuclear systems, which are interesting for the
$\beta\beta$ decay. Only 
for the double beta decay candidate $^{48}Ca$  this has been found possible
 by considering the
full (pf) shell \cite{ret95}. 
In the heavier nuclei the number of basis 
states increases astronomically and it is difficult to perform 
a realistic shell-model calculation without very severe truncations. 

In earlier studies the weak coupling scheme has been employed, 
starting from a product of neutron and proton wave functions and employing
truncations according the energies of the unpertubated proton and 
neutron states \cite{hax84,sin88,gru85}. 
Nowadays, there are some shell model codes 
\cite{cau96,zha90,john97,nak96} 
which can treat some of the heavier nuclear systems 
undergoing double beta decay in a considerably more realistic way. 
We remind also that the feasibility of the shell model 
calculations is growing
with increasing computer facilities. 

The shell model calculations generally have resulted 
in larger transition matrix elements than the QRPA. 
One may surmise that the stringent truncation imposed on shell model 
spaces exclude configurations responsible for the reduction
by destructive interference. 
This idea is supported by the shell model studies of the
$2\nu\beta\beta$-decay of $^{48}Ca$ \cite{zha90} and $^{92}Mo$ 
\cite{john97}, in which the dependence of 
the $2\nu\beta\beta$-decay matrix elements on the configuration
space used have been examined.
Another reason can be the omission of the spin-orbit 
partners. Recently, it has been found that there is a  destructive 
interference between  "spin-flip" and "non spin-flip" contributions 
to the $2\nu\beta\beta$-decay transition leading to
the suppression of $2\nu\beta\beta$-decay matrix elements
\cite{rad95,civn95}. In addition the matrix elements obtained 
within the shell model can be suppressed
by the renormalization of the Gamow-Teller operators in order 
to account for about $20\%$ 
quenching of the beta decay strength.
Till now, the origin of this quenching phenomenon
is not well understood and can be explained by a number of reasons,
e.g.  by influence of many-particle configurations or by the
 renormalization of axial vector and vector weak interactions constants 
due to nuclear medium effects leading to $g_A/g_V\approx 1$.

\subsection{QRPA and renormalized QRPA}

The pn-QRPA has been found  to be a powerful model, 
considering its simplicity, to describe nuclear matrix elements 
in ordinary beta decay. In addition, the QRPA can handle of 
great number of intermediate states. Therefore,
the pn-QRPA has been the most frequently used nuclear structure method
to deal with the nuclear structure aspects of the $\beta\beta$ decay
for open shell systems [see e.g. 
\cite{vog86,civ87,mut88,pan92,krm94,tom87,eng88,mut89}]. 
In the double beta decay the initial nucleus decays to final nucleus 
through virtual excitations of all states of the intermediate nucleus.
The excited states are obtained by solving the well-known QRPA equations,
which yield all excited states of a given 
even-even nucleus. The advantage of the QRPA calculation in respect to the
shell model approach is that it can be performed in large model spaces. 
The QRPA produces a considerably smaller number of states than the 
shell model as it is based on a harmonic approximation. 

A shortcoming  of the pn-QRPA is that it
frequently overestimates  the ground state
correlations, leading to instabilities and finally to collapse
of the QRPA solution. The collapse is caused by
generation of too many correlations with increasing 
strength of the attractive proton-neutron interaction. Unfortunately,
the physical value of this force is usually close to the point in which
the QRPA solutions collapse. This behavior has its origin in the quasi-boson
approximation (QBA)
violating the Pauli exclusion principle and causing the QRPA excitation 
operators behave like bosons \cite{toi95,simn96}. The increasing
violation of the Pauli principle is generated by  
excessive ground state correlations. To overcome this 
difficulty the renormalized QRPA 
have been proposed \cite{toi95,simn96},
which take into account the Pauli exclusion principle in an 
approximate way.  The price paid for bypassing the collapse 
in the RQRPA is the violation of the Ikeda sum rule \cite{krm96,hirm96}.

The pn-QRPA \cite{vog86,civ87,mut88}, full-QRPA \cite{cheo93}, 
pn-RQRPA \cite{toi95} and full-RQRPA \cite{simn96} formalisms are fairly 
well known. So we briefly  mention only its main features here. 

The QRPA as well as the RQRPA formalisms consist of two main steps: (i) The
Bogoliubov transformation smears out the nuclear Fermi surface 
over a relatively large number of orbitals 
(ii)  The equation of motion in the quasiparticle basis determines then the
excited states. 

If the proton-proton, neutron-neutron
and proton-neutron pairing is considered,
the particle ($c^{+}_{\tau a m_{a}}$ and
$c^{}_{\tau a m_{a}}$, $\tau = p,n$) and quasiparticle 
($a^{+}_{\mu a m_{a}}$ and $a^{}_{\mu a m_{a}}$, $\mu = 1,2$) 
creation and annihilation operators for
spherical shell model states (The label $a$ designates quantum numbers 
$n^{}_a, l^{}_{a}, j^{}_{a}$.) are related each to other by the 
Hartree-Fock-Bogoliubov (HFB) transformation: 
\begin{equation}  
\left( \matrix{ c^{+}_{p k m_{k} } \cr
c^{+}_{n k m_{k}} \cr {\tilde{c}}_{p k m_{k} } \cr 
{\tilde{c}}_{n k {\tilde{m}}_{k}} 
}\right) = \left( \matrix{ 
u_{k 1 p} & u_{k 2 p} & -v_{k 1 p} & -v_{k 2 p} \cr 
u_{k 1 n} & u_{k 2 n} & -v_{k 1 n} & -v_{k 2 n} \cr
v_{k 1 p} & v_{k 2 p} & u_{k 1 p} & u_{k 2 p} \cr  
v_{k 1 n} & v_{k 2 n} & u_{k 1 n} & u_{k 2 n} }\right)
\left( \matrix{ a^{+}_{1 k m_{k}} \cr
a^{+}_{2 k m_{k}} \cr {\tilde{a}}_{1 k m_{k}} \cr 
{\tilde{a}}_{2 k m_{k}} }\right).
\label{qrpa.1}  
\end{equation} 
where the tilde $\sim$ indicates  time reversal 
($a_{\tau a {\tilde m}_{a}}$ = $(-1)^{j_{a} - m_{a}}a^{}_{\tau a -m_{a}}$). 
The occupation amplitudes $u$ and 
$v$  and the single quasiparticle energies $E^{}_{a \alpha}$ 
are obtained by solving the adequate HFB equation \cite{cheo93}.  
Since the  model space considered is  finite, the pairing
interactions are renormalized by the strength parameters
$d_{pp}$, $d_{nn}$ and $d_{pn}$ \cite{cheo93}
to the empirical gaps defined by Moeller and Nix
\cite{mn92}. The numerical values 
of the $d_{pp}$ and $d_{nn}$ are close to unity.
A $d_{pn}$ value higher than unity is the price
paid for the spherical symmetry of the model which excludes the 
treatment of the T=0 pairing. The J=0 T=0 pairs can be treated
in a BCS or even HFB approach only due to deformation. The T=0 
pairing is effectively taken into account by the renormalization 
of the T=1 J=0 n-p interaction leading to a higher value of $d_{pn}$. 
For large enough model space the value of $d_{pn}$ is about 1.5,
what is reasonable \cite{simf97}. 

In the limit without proton-neutron pairing one has: 
$u^{}_{2p}$ = $v^{}_{2p}$ = $u^{}_{1n}$ = $v^{}_{1n}$ = 0. In this case
the Bogoliubov transformation in Eq. (\ref{qrpa.1}) is reduced  to two 
BCS transformations,  for protons ($u^{}_{1p}=u^{}_p$, 
$v^{}_{1p}=v^{}_p$) and  for neutrons 
($u^{}_{2n}=u^{}_n$, $v^{}_{2n}=v^{}_n$) separately.

In the framework of the full-QRPA or full-RQRPA
the $m^{th}$ excited states with angular momentum $J$ and projection $M$ 
is created by a phonon-operator $Q$ with the properties
\begin{equation}
Q^{m\dagger}_{JM}|0^+_{RPA}\rangle
=|m,JM\rangle \qquad \mbox{and} \qquad
Q|0^+_{RPA}\rangle=0.
\label{qrpa.2}  
\end{equation}
Here, $|0^+_{RPA}\rangle$ is the ground state of the initial or 
the final nucleus.
The phonon-operator $Q$ takes the following form
\begin{eqnarray}
Q^{m\dagger}_{JM^\pi}&=&\sum_{k l}
  X^m_{12}(k,l,J) A^{\dagger}_{12}(k,l,J,M)
+ Y^m_{12}(k,l,J) {\tilde{A}}_{12}(k,l,J,M)+\nonumber \\
&&\sum_{k \leq l \atop \mu = 1,2}
  X^m_{\mu\mu}(k,l,J) A^{\dagger}_{\mu\mu}(k,l,J,M)
+ Y^m_{\mu\mu}(k,l,J) {\tilde{A}}_{\mu\mu}(k,l,J,M).
\label{qrpa.3}  
\end{eqnarray}
$ A^{\dagger}_{\mu\nu}(k,l,J,M)$ is two quasiparticle 
creation and annihilation operator coupled
to angular momentum $J$ with projection $M$ namely
\begin{eqnarray}
A^{\dagger}_{\mu\nu}(k,l,J,M) 
&=& n(k\mu, l\nu) \sum^{}_{m_k , m_l }
C^{J M}_{j_k m_k j_l m_l } a^\dagger_{\mu k m_k} a^\dagger_{\nu l m_l}\,,
\nonumber \\
n(k\mu, l \nu)&=&
(1+(-1)^J\delta_{kl}\delta_{\mu \nu})/(1+\delta_{kl}
\delta_{\mu \nu})^{3/2}.
\label{qrpa.4}  
\end{eqnarray}
We note that if proton-neutron pairing is neglected, the phonon 
operator in Eq. (\ref{qrpa.3}) decouples on two phonon operators,
one for charge changing and second for non-charge changing modes
of nuclear excitations.

In the full-RQRPA the commutator
of two bifermion operators is replaced with its expectation value in
the correlated QRPA ground state $|0^+_{QRPA}>$
(renormalized quasiboson approximation). We have
\begin{eqnarray}
&& \big<0^+_{QRPA}\big|\big
[A^{}_{\mu \nu}(k, l, J, M),A^+_{\mu' \nu'}(k', l', J, M)\big]
\big|0^+_{QRPA}\big> =
\nonumber \\ 
&&n(k\mu, l\nu) n(k'\mu', l'\nu') 
\Big( \delta_{kk'}\delta_{\mu \mu' }\delta_{ll'}
\delta_{\nu\nu'} -
\delta_{lk'}\delta_{\nu \mu'}\delta_{kl'}
\delta_{\mu \nu'}(-1)^{j_{k}+j_{l}-J}\Big)\times \nonumber \\
\lefteqn{\underbrace{
\Big\{1
\,-\,\frac{1}{\hat{\jmath}_{l}}
<0^+_{QRPA}|[a^+_{\nu l}{\tilde{a}}_{\nu l}]_{00}|0^+_{QRPA}>
\,-\,\frac{1}{\hat{\jmath}_{k}}
<0^+_{QRPA}|[a^+_{\mu k}{\tilde{a}}_{\mu \tilde{k}}]_{00}|0^+_{QRPA}>
\Big\}
}_{
=:\displaystyle {\cal D}_{\mu k, \nu  l; J^\pi}
},} &&
\nonumber \\
\label{qrpa.5}  
\end{eqnarray}
with  $\hat{\jmath}_k=\sqrt{2j_k+1}$.
If we replace $|0^+_{QRPA}>$ in Eq. (\ref{qrpa.5}) with the uncorrelated
HFB ground state, we obtain the quasiboson approximation (i.e. 
${\cal D}_{\mu k, \nu  l; J^\pi}=1$), which violates the Pauli 
exclusion principle by neglecting the 
terms coming from the Fermion commutation
rules of the quasi-particles.  
The QBA assumes that  pairs of quasiparticles 
obey the commutation relations of bosons.

The full-RQRPA takes into account the Pauli exclusion
principle more carefully. From Eqs. (\ref{qrpa.2}) and (\ref{qrpa.3})
one can derive the RQRPA equation
\begin{equation}
  \underbrace{{\cal D}^{-1/2}\left(
    \begin{array}{cc}
       \cal A &\cal B\\
       \cal -B &\cal -A
    \end{array}
    \right)
    {\cal D}^{-1/2}}
    _{ \textstyle \overline{\cal A},\overline{\cal B}}
    \;
    \underbrace{{\cal D}^{1/2}\left(
    \begin{array}{c}
       X^m\\
       Y^m
    \end{array}
    \right)}_{\textstyle \overline{X}^m, \overline{Y}^m}
    = \Omega^m_{J^\pi}
    \underbrace{{\cal D}^{1/2}
    \left(
    \begin{array}{c}
       X^m\\
       Y^m
    \end{array}
    \right)}_{\textstyle\overline{X}^m, \overline{Y}^m}\, .
\label{qrpa.6}  
\end{equation}
The matrices $\cal{A}$ and
$\cal{B}$  are given as follows:
\begin{eqnarray}
\label{qrpa.7}
{\cal A}^{a\alpha b\beta}_{k\mu l\mu,J} &=&
\left\langle 0^+_{RPA}\right|\left[
A_{\alpha\beta}(a,b,J,M)
,\left[H,A^{\dagger}_{\mu\nu}(k,l,J,M)
\right]\right]\left|0^+_{RPA}\right\rangle\,,\\
\label{qrpa.8}
{\cal B}^{a\alpha b\beta}_{k\mu l\mu,J} &=&
\left\langle 0^+_{RPA}\right|\left[
A_{\alpha\beta}(a,b,J,M),\left[
H,{\tilde{A}}_{\mu\nu}(k,l,J,M)\right]\right]\left|0^+_{RPA}\right\rangle\,,
\end{eqnarray}
where $H$ is the quasiparticle nuclear Hamiltonian \cite{simn96}.
The coefficients 
${\cal D}_{\mu k, \nu  l; J^\pi}$ 
are determined by solving numerically the system of 
equations\cite{toi95,simn96}:
\begin{eqnarray}
{\cal D}_{k\mu l\nu J^\pi} &=& 1-\frac{1}{\hat{\jmath}_k^2} 
\sum_{k'\mu' \atop J'^{\pi'} m}
{\cal D}_{k\mu k'\mu'J'^{\pi'}}\hat{J}'^2 \big|
{\overline{Y}}^m_{\mu \mu'}(k, k', J'^{\pi'})  \big|^2
\nonumber \\
&& ~~~~~-\frac{1}{\hat{\jmath}_l^2}\sum_{l'\nu' \atop J'^{\pi'} m}
{\cal D}_{l\nu
l'\nu'J'^{\pi'}}\hat{J}'^2 \big |{\overline{Y}}^m_{\nu \nu'}(l, l', J'^{\pi'})
  \big|^2 .
\label{qrpa.9}  
\end{eqnarray}
The selfconsistent scheme of the calculation of the forward-
(backward-) going free variational amplitude   
${\overline{X}}^m_{}$ (${\overline{Y}}^m_{}$), 
the energies of the excited states 
$\Omega^m_{J^\pi}$ for every multipolarity $J^\pi$
and the coefficients ${\cal D}_{\mu k, \nu  l; J^\pi}$ 
is a double iterative problem which requires the solution of 
coupled Eqs. (\ref{qrpa.6}) and (\ref{qrpa.9}).

In order to calculate
double beta transitions two fully independent RQRPA calculations are
needed, one to describe the beta transition from the initial to the
intermediate nucleus and another one for the beta transition from the
intermediate to the final nucleus. For that purpose the one-body transition
densities of the charge changing operator has to be evaluated. They take
the following form: 
\begin{eqnarray}
\label{qrpa.10}   
<J^\pi m_i \parallel [c^+_{pk}{\tilde{c}}_{nl}]_J \parallel 0^+_i>=
\sqrt{2J+1}\sum_{\mu,\nu = 1 , 2} m(\mu k,\nu l)\times \\ \nonumber
\left [
u_{k \mu p}^{(i)} v_{l \nu n}^{(i)} {\overline{X}}^{m_i}_{\mu\nu}(k,l,J^\pi)
+v_{k \mu p}^{(i)} u_{l \nu n}^{(i)} {\overline{Y}}^{m_i}_{\mu\nu}(k,l,J^\pi)
\right ]
\sqrt{{\cal D}^{(i)}_{k \mu l \nu l J^\pi}},
\end{eqnarray}
\begin{eqnarray}
\label{qrpa.11}   
<0_f^+ \parallel \widetilde{ [c^+_{pk'}{\tilde{c}}_{nl'}]_J} 
\parallel J^\pi m_f>=
\sqrt{2J+1}\sum_{\mu,\nu=1,2}m(\mu k',\nu l')\times \\ \nonumber 
\left [
v_{k' \mu p}^{(f)} u_{l' \nu n}^{(f)} 
{\overline{X}}^{m_f}_{\mu\nu}(k',l',J^\pi)
+u_{k' \mu p}^{(f)} v_{l' \nu n}^{(f)} 
{\overline{Y}}^{m_f}_{\mu\nu}(k',l',J^\pi)
\right ]
\sqrt{{\cal D}^{(f)}_{k' \mu l' \nu J^\pi}},
\end{eqnarray}
with 
$m(\mu a,\nu b)=\frac{1+(-1)^J\delta_{\mu \nu}\delta_{ab}}{(1+\delta_{\mu\nu}
\delta_{ab})^{1/2}}$. We note that the  
${\overline{X}}^{m}_{\mu\nu}(k,l,J^\pi)$ and
${\overline{Y}}^{m}_{\mu\nu}(k,l,J^\pi)$ amplitudes are calculated 
by the renormalized QRPA equation only for the configurations 
$\mu a \leq \nu b$ ( i.e., if 
$\mu = \nu$ the orbitals a and b are ordered $a \leq b$, and 
if $\mu = 1$, $\nu = 2$, the orbitals need not to be  ordered).
For the amplitudes 
${\overline{X}}^{m}_{\mu\nu}(k,l,J^\pi)$ and
${\overline{Y}}^{m}_{\mu\nu}(k,l,J^\pi)$ in Eqs. 
(\ref{qrpa.10}) and (\ref{qrpa.11}) exist the following symmetry relations:
\begin{equation}
{\overline{X}}^{{m}}_{\mu \nu}(k,l,J^\pi)
 = -(-1)^{{{j}}_{{k}}+{{j}}_{{l}}-{J}}
{\overline{X}}^{{m}}_{\nu \mu}(l,k,J^\pi), ~~~
{\overline{Y}}^{{m}}_{\mu \nu}(k,l,J^\pi)
 = -(-1)^{{{j}}_{{k}}+{{j}}_{{l}}-{J}}
{\overline{Y}}^{{m}}_{\nu \mu}(l,k,J^\pi). 
\label{qrpa.12}    
\end{equation} 
The index i (f) indicates that the quasiparticles and the excited
states of the nucleus are defined with respect to the initial (final)
nuclear ground state $|0^+_i>$ ($|0^+_f>$). 
We note that for ${\cal D}_{k \mu l \nu J^\pi}=1$ 
(i.e. there is no renormalization) and $u_{{2p}} =
\upsilon_{{2p}} = u_{{1n}} = \upsilon_{{1n}} = 0$ (i.e.
there is no proton-neutron pairing), Eqs.\ (\ref{qrpa.10}) and
(\ref{qrpa.11}) reduce to the expressions of the standard 
pn-QRPA.

The two sets of intermediate nuclear states generated from the 
initial and final ground states are not identical in the
 QBA or RQBA schemes considered. Therefore the overlap factor 
of these states is introduced in the theory as follows\cite{sims97}:
\begin{eqnarray}
\label{qrpa.13}    
<J_{m^{}_{f}}^+ | J_{m^{}_{i}}^+> \approx 
[Q^{ m_f}_{JM}, Q^{+ m_i}_{JM}] \approx 
\sum_{\mu k \leq \nu l,~~ \mu' k' \leq \nu' l'}
\delta_{kk'}\delta_{ll'}{\tilde{u}}_{k\mu\mu'} {\tilde{u}}_{l\nu\nu'} 
\times \\ \nonumber
\big(
\overline{X}^{m_{i}^{}}_{\mu \nu}(k, l, J)
\overline{X}^{m_{f}^{}}_{\mu' \nu'}(k, l, J)-
\overline{Y}^{m_{i}^{}}_{\mu \nu}(k, l, J)
\overline{Y}^{m_{f}^{}}_{\mu' \nu'}(k, l, J) 
\big),
\end{eqnarray}
with 
\begin{equation}
{\tilde{u}}_{k \mu \mu'} = u^{(i)}_{k\mu}u^{(f)}_{k\mu'} +
v^{(i)}_{k\mu}v^{(f)}_{k\mu'}.
\label{qrpa.14}    
\end{equation}
Here, $Q^{ m_f}_{JM}$  and $Q^{+ m_i}_{JM}$ are respectively
phonon annihilation and creation operators for the initial 
and final nuclear states. We note that in the previous calculations
$\tilde{u}$ was approximated by  unity. However, in that case
the overlap factor is dependent on the phases of the 
BCS/HFB occupation  amplitudes u's and v's, which are in principal arbitrary.
A negligible difference between the results with the above overlap
containing $\tilde{u}$ and with the overlap with $\tilde{u}=1$
one obtains only if the phases of the BCS/HFB amplitudes are
chosen so that $\tilde{u}$ is positive for each level.

The double beta decay nuclear matrix elements are usually 
transformed to ones containing two-body matrix elements 
by using the second quantization formalism. One arrives
at the expression: 
\begin{eqnarray}
<O_{12}>=
\sum_{{k l \acute{k} \acute{l} } \atop {J^{\pi}
m_i m_f {\cal J}  }}
~(-)^{j_{l}+j_{k'}+J+{\cal J}}(2{\cal J}+1)
\left\{
\begin{array}{ccc}
j_k &j_l &J\\
j_{l'}&j_{k'}&{\cal J}
\end{array}
\right\}\times ~~~~~\nonumber \\
< 0_f^+ \parallel 
\widetilde{[c^+_{pk'}{\tilde{c}}_{nl'}]_J} \parallel J^\pi m_f>
<J^\pi m_f|J^\pi m_i>
<J^\pi m_i \parallel [c^+_{pk}{\tilde{c}}_{nl}]_J \parallel 
0^+_i >\nonumber \\
\times<pk,pk';{\cal J}|f(r_{12})\tau_1^+ \tau_2^+ {\cal O}_{12}
f(r_{12})|nl,nl';{\cal J}>.~~~~~~~~~~
\label{qrpa.15}    
\end{eqnarray}
with short-range correlation functions
\begin{equation}
f(r)=1-e^{-\alpha r^2 }(1-b r^2) \quad \mbox{with} \quad
\alpha=1.1 \mbox{fm}^2 \quad \mbox{and} \quad  b=0.68 \mbox{fm}^2,
\label{qrpa.16}    
\end{equation}
which takes into account the short range repulsion of the nucleons.
In the case of the nuclear matrix element $M^{2\nu}_{GT}$ in Eq. (\ref{two.8})
${\cal O}_{12}$ is replaced by 
$2 {\mathbf{\sigma}}_1 \cdot {\mathbf{\sigma}}_2 
/(\Omega^{m_i}_{1^+}+\Omega^{m_f}_{1^+})$ and in the case of
neutrinoless double beta decay matrix elements in Eqs. (\ref{neu.9})
and (\ref{susy.8}) ${\cal O}_{12}$ is replaced by 
${\cal O}_{I}({\mathbf{r}}_1,{\mathbf{r}}_2,
{\mathbf{\sigma}}_1,{\mathbf{\sigma}}_2) $ [see Table 
\ref{tabneu.1}]. 

In the QRPA and  RQRPA calculations it is necessary to introduce 
the renormalization parameters $g_{pp}$ and $g_{ph}$ for particle-particle
and particle-hole channels of the G-matrix interaction of 
the nuclear Hamiltonian $H$, which in principle should be close to unity. 
Slight deviations from unity can be expected because the full Hilbert 
space can 
not be taken into account.  
The $g_{ph}$ value is usually fixed by physical arguments, e.g. 
by fitting the position of giant Gamow-Teller resonance.
The $g_{pp}$ is considered as free parameter  
and the dependence of the matrix element on $g_{pp}$ is carried out. 
The problem regarding $g_{pp}$ is crucial to predictability of the
QRPA calculation. It has been shown that $\beta^+$ transitions 
(second branch of
the double beta decay in reverse) 
to low-lying states are very sensitive to $g_{pp}$.
The introduction of $g_{pp}$ as a phenomenological factor makes it possible 
to reproduce the observed $2\nu\beta\beta$-decay half-lifes.

\subsection{Operator expansion method}

The Operator expansion method 
(OEM) is a nuclear structure method for the $2\nu\beta\beta$-decay,
which has the advantage to avoid the explicit sum over the intermediate
nuclear states. 

There are two different ways to derive 
the $2\nu\beta\beta$-decay OEM transition operators. 
In the first  approach  (OEM1) \cite{chi89} the expansion of the
denominators of $M^{2\nu}_{GT}$ in Eq.\ (\ref{two.8}) in  Taylor series
is used. In the second approach proposed (OEM2)
\cite{sim89} the OEM is derived from the integral representation of 
the nuclear matrix element 
$M^{2\nu}_{GT}$ in Eqs.\ (\ref{two.6}) and (\ref{two.7}). 
It has been found that the OEM2 offers advantages over OEM1 as there 
are no problems with convergence  of the power series expansion
of the denominator \cite{simm97,sims97}, 
which has been a subject of criticism \cite{eng92}. 
In addition, the approximations under consideration are better controlled. 

The OEM is based on two main assumptions \cite{chi89,sim89}:
i) It is assumed that the kinetic energy operator can be ignored in the
resulting commutators [ see Eqs.\ (\ref{two.6}) and (\ref{two.7}) ]
and therefore the nuclear Hamiltonian H is
represented only by two-body interaction terms.
ii) Only two-body terms are retained in evaluating each commutator and
higher order terms are neglected. These approximations are
supported by the following arguments:  First, if we 
consider only the one-body  part of the nuclear Hamiltonian $H_0$,
the $2\nu\beta\beta$-decay transition operator [A(t/2), A(-t/2)] 
in (\ref{two.6}) is an one body operator \cite{simm97,sims97}. As the 
$2\nu\beta\beta$-decay operator should be at least a two-body operator 
to change two neutrons into two protons, it means that the one-body
operator of H plays a less important role in the evaluation of 
$M^{2\nu}_{GT}$. Second, in the case of the QBA 
the commutator [A(t/2), A(-t/2)] is just a constant but within the
OEM it is a two-body operator. It means that the OEM approximations
go beyond the QBA and RQBA \cite{simm97,sims97}.

Further, it has been shown  
that the energy  difference between the initial and final nuclear states 
plays an important role in the derivation of the OEM transition operators 
and that the consideration of Coulomb interaction is crucial 
for the OEM \cite{sim90}. If the approximate nuclear two-body Hamiltonian
$H$ does not contain a Coulomb interaction term \cite{chi89,sim89,wu91} or
the Coulomb interaction is considered in the way  of refs. 
\cite{gmi90,mut93,kad95},
the derivation of the  OEM transition operators is not consistent. 
Therefore, an effective Coulomb interaction term $V_{C}$ has to be
considered, which 
represents partially the  one body terms of the nuclear Hamiltonian
\cite{sim90,simm97,sims97}. 
The approximate OEM two-body nuclear Hamiltonian H containing $V_{C}$,
central $V_{CN}$ and tensor $V_{TN}$ interactions is given as follows
(the notation of ref. \cite{mut93} is used):
\begin{eqnarray}
H \approx V_C + V_{CN} +V_{TN},
\label{oem.1}
\end{eqnarray}
where
\begin{eqnarray}
V_C&=&\frac{1}{2}\sum_{i\neq j} ~(E_f -E_i ) ~O^\tau_{ij},
~~~~~~O^\tau_{ij}=\frac{1}{4}(1+\tau^0_i )(1+\tau^0_j ),\\
V_{CN}&=&\frac{1}{2}\sum_{i\neq j} 
~[~~   (~ g_{SE}(r_{ij})~\Pi^r_e(ij) ~+~
      g_{SO}(r_{ij})~\Pi^r_o(ij)~ )~\Pi^\sigma_s(ij) +
\nonumber \\
&&\phantom{~~~\frac{1}{2}\sum_{i\neq j}}
    ( ~g_{TE}(r_{ij})~\Pi^r_e(ij) ~+~
      g_{TO}(r_{ij})~\Pi^r_o(ij)~ ) ~\Pi^\sigma_t(ij) ~~],
\\
V_{TN}&=&\frac{1}{2}\sum_{i\neq j} 
~(~ g_{TNE}(r_{ij})~\Pi^r_e(ij) + 
g_{TNO}(r_{ij} )~\Pi^r_o(ij)~)~S_{ij}.
\label{oem.2}
\end{eqnarray}

In the framework of the OEM2 for the nuclear matrix element $M^{2\nu}_{GT}$
one obtains \cite{simm97,sims97}:
\begin{eqnarray}
M^{2\nu}_{GT}&=&<0^+_f|\frac{1}{2} {\cal{P}}
\sum \limits_{i \ne j} \tau^+_i \tau^+_j
~(~{\cal V}^{singlet}(r_{ij})~\Pi^\sigma_s(ij) ~+ \nonumber \\
&&\phantom{ \sum \limits_{i \ne j} \tau^+_i \tau^+_j }
 {\cal V}^{triplet}(r_{ij})~\Pi^\sigma_t(ij) ~+~
 {\cal V}^{tensor}(r_{ij})~S_{ij}~)~|0^+_i>,
\label{oem.3}
\end{eqnarray}
where, 
\begin{eqnarray}
{\cal V}^{singlet}&=&\frac{-2}{g_{TE}-g_{SE}-4g_{TNE}+\Delta } -
 \frac{4}{g_{TE}-g_{SE}+2g_{TNE}+ \Delta } \\
{\cal V}^{triplet}&=&\frac{1}{3} [ 
\frac{4}{\Delta} +\frac{4}{-6g_{TNO}+\Delta}+
\frac{4}{6g_{TNO}+\Delta}  -  \nonumber \\
&& \frac{2}{g_{SO}-g_{TO}+4g_{TNO}+\Delta } -
\frac{4}{g_{SO}-g_{TO}-2g_{TNO}+\Delta } ], \\
{\cal V}^{tensor}&=& \frac{1}{3} [ \frac{1}{\Delta}+
\frac{1}{-6g_{TNO}+\Delta } -
\frac{2}{6g_{TNO}+\Delta } + \nonumber \\
&&\frac{1}{g_{SO}-g_{TO}+4g_{TNO}+\Delta } -
\frac{1}{g_{SO}-g_{TO}-2g_{TNO}+\Delta } ].
\label{oem.4}
\end{eqnarray}
Here ${\cal{P}}$ denotes the Principle value integration.
We note that the central and tensor interactions appear only in the 
denominators of the OEM potentials. As $\Delta$ 
(half energy release for $2\nu\beta\beta$-decay process) is 
about few MeV, only the long range part of nucleon-nucleon 
potential is relevant for the calculation. Therefore, the OEM
potentials are not expected to be sensitive to the type of realistic
effective interaction, e.g. Bonn or Paris ones.

For the calculation of $M^{2\nu}_{GT}$ in Eqs. (\ref{oem.3}) and
(\ref{oem.4}) it is
necessary to know the wave functions of the initial and final nucleus.
The OEM can be combined with the ground state wave functions 
of the QRPA or renormalized RQRPA models (OEM+RPA)
\cite{wu91,mut93,ves98}. Recently, the consistently derived
OEM transition operators have been combined with pn-RQRPA wave
functions of the initial and final nuclear systems to calculate
$2\nu\beta\beta$-decay of $^{76}Ge$ \cite{ves98}. 
A strong dependence of the matrix element $M^{2\nu}_{GT}$
has been found on the  choice of the pn-RQRPA ground state
wave functions. Therefore, there is an interest to combine 
OEM-transition operators in Eqs. (\ref{oem.3}) and (\ref{oem.4}) 
with other ground state wave functions, e.g. shell model ones.

\subsection{Alternative methods}

Till now no consistent many body approach is available for the
calculation of the many-body Green function governing
the $2\nu\beta\beta$-decay process, because of the computational 
complexity of the problem. The small predictive power of the QRPA 
is the motivation that stimulated us to seek an
alternative way, which might produce more reliable results.

The most promising are those methods, which can guarantee the particle
number conservation and an explicit consideration of the Pauli 
principle. The two-vacua RPA 
(TVRPA) fulfill both criteria \cite{ten95}. The trial 
nuclear wave functions of the intermediate odd-odd nucleus are 
constructed as linear combinations of proton-particle neutron-hole
excitations from the ground state of the initial even-even (N,Z)
nucleus and at the same time, as neutron-particle proton-hole excitations
from the ground state of the final (N-2,Z+2) nucleus. Till now,
the method has been 
applied only to the $2\nu\beta\beta$-decay of $^{48}Ca$. In order to understand
its predictive power, it is clearly necessary to perform 
more calculations of the beta strength distributions and double beta decay
matrix elements.

In the past few years a new
method for the solution of the nuclear many-body problem
has been developed, which uses 
the  shell model Monte Carlo technique (SMMC) \cite{koo97}.
The SMMC method enforces the Pauli principle exactly and
scales more gently with the problem of  the single particle basis 
size than the traditional shell model direct-diagonalization
approach. The SMMC has been found successful in description of the 
Gamow-Teller strength distributions and  has been also
applied for the $2\nu\beta\beta$-decay of $^{48}Ca$ (complete $pf$ shell)
and $^{76}Ge$ ($0f_{5/2}$,$1p$,$0g_{9/2}$ model space) giving promising
results \cite{koo97}. However, the dependence of $M^{2\nu}_{GT}$ on effective 
interaction and single particle energies remains to be investigated.

Other approaches looking for the physical cause of the suppression
mechanism of the $2\nu\beta\beta$-decay are 
 based on the approximate Wigner 
spin-isospin $SU(4)_{\sigma\tau}$ 
symmetry in nuclei \cite{ber90,rum95}. The possibility 
of using $SU(4)_{\sigma\tau}$ symmetry to analyze nuclear excitations
(Gamow-Teller and isobaric analog resonances) has been discussed e.g. 
in \cite{vlad92}. We know that $SU(4)_{\sigma\tau}$ is not an exact 
symmetry and is broken by the spin-orbit and Coulomb forces. It is 
worthwhile to notice that
if the exact $SU(4)_{\sigma\tau}$ symmetry is considered,
$2\nu\beta\beta$-decay is forbidden \cite{moe94}, 
i.e. $2\nu\beta\beta$-decay is
determined by the mechanism responsible for  breaking of this symmetry.
In the approach proposed in \cite{rum95} 
the spin-orbit part of the mean field of the nucleus was assumed to be
responsible for the symmetry breaking.

\section{Results and Discussion}

\subsection{QRPA and violation of the Pauli principle}

Most of the  calculations of the $2\nu\beta\beta$-decay matrix elements
for nuclei heavier than $^{48}Ca$ rely on  QRPA. However, 
the calculated matrix elements are uncertain because of their 
great sensitivity to $J=1^+$, $T=0$ particle-particle component 
of the residual interaction governed by the $g_{pp}$ parameter
\cite{vog86,civ87,mut88}. The $2\nu\beta\beta$-decay matrix element
$M^{2\nu}_{GT}$
is reduced by the enhanced ground state correlations with increasing 
strength of proton-neutron interaction and it even crosses zero.
The dependence of $M^{2\nu}_{GT}$ on $g_{pp}$
for the $2\nu\beta\beta$-decay of $^{76}Ge$ 
calculated within the pn-QRPA for three different model spaces 
is presented in the Fig. (\ref{res.1})a \cite{sims97}. For 
9, 12 and 21 level  model spaces the collapse of the pn-QRPA occurs
near the  critical interaction strength given by $g_{pp}$
values 1.3, 1,1 and 0.9, respectively.  A larger model space means more 
ground state correlations, i.e. a collapse of the pn-QRPA solution
for smaller $g_{pp}$. 

It was long believed that 
this curious behavior of the $2\nu\beta\beta$-decay 
matrix elements in the QRPA is typical only for this decay mode. 
Recently, it has been found 
that in the framework of the pn-QRPA for a large enough model space
the $0\nu\beta\beta$-decay matrix elements shows similar
instability with respect to $g_{pp}$ as the one known from 
the $2\nu\beta\beta$-decay mode. The value of the matrix
element crosses zero and it is then difficult to  make definite
predictions about the transition rate [see Fig. (\ref{res.2})a]. 
This fact is not surprising
as the origin of this dependence is the product of two
one body transition densities of the charge changing operator
to the lowest (or one of the lowest)  
$1^+$ state of the intermediate nucleus namely
\begin{equation}
< 0_f^+ \parallel 
\widetilde{[c^+_{pk'}{\tilde{c}}_{nl'}]_1} \parallel 1^+_{1-f}>
<1^+_{1-f}|1^+_{1-i}>
<1^+_{1-i} \parallel [c^+_{pk}{\tilde{c}}_{nl}]_J \parallel 
0^+_i >, ~~(m_i=m_f=1),
\end{equation}
which is present in the calculation of each the double beta
decay matrix element via the QRPA and its extensions (see 
Eqs. [\ref{qrpa.16}]) except the simple double Fermi transition.
The theoretical nuclear matrix elements 
show large uncertainties, since they depend on the  
 ground state correlations, which are overestimated in QRPA. 

It is worthwhile to notice that
there is no mechanism in  QRPA, which  prevents the increase 
of the ground state correlation to unreasonable large values 
with increasing $g_{pp}$,  leading finally to a collapse 
of QRPA solution.
The investigation of the $M^{2\nu}_{GT}$
within the pn-RQRPA \cite{toi95} and the full-RQRPA \cite{simn96}
has shown 
that the main source of the ground state instability in  QRPA
is the  violation of the Pauli exclusion principle, if one has large
 ground state correlations.

The RQRPA, which considers the  Pauli exclusion principle in an approximate
way, shifts  the collapse of the QRPA outside the physical range of 
$g_{pp}$ and  shows a  less sensitive dependence of $M^{2\nu}_{GT}$ on 
$g_{pp}$  \cite{toi95,simn96,krm96,muto97}. 
Fig. (\ref{res.1})b shows 
$M^{2\nu}_{GT}$ for  the $2\nu\beta\beta$-decay of $^{76}Ge$
calculated by pn-RQRPA  \cite{sims97}. 
We see that the crossing point through zero is shifted to  larger $g_{pp}$ 
in comparison with Fig. (\ref{res.1})a. A  steeper slope of 
$M^{2\nu}_{GT}$ as a function of $g_{pp}$ 
for larger model spaces may indicate that the approximate
treatment of the Pauli exclusion principle by the pn-RQRPA
is for larger $g_{pp}$ not good enough.

The RQRPA offers considerable less sensitive results to $g_{pp}$ 
in comparison with the QRPA also  for the $2\nu\beta\beta$-decay
transitions to the $0^+$ and $2^+$ excited states of the final
nucleus \cite{sch97,bara97}. The same is valid also for the calculation
of $M^{0\nu}_{mass}=M^{0\nu}_{GT}(1-\chi_F)$ 
associated with the neutrino mass mechanism of the
$0\nu\beta\beta$-decay [see Fig. (\ref{res.2})] \cite{simf97,ves97}. 

In both QRPA and RQRPA treatments of the double beta decay matrix elements
one has  two sets for the same  intermediate nuclear states generated from 
initial and final nuclei, which do not coincide with each other. 
The overlap factor of these two sets is introduced artificially and
the effect of this inaccuracy is not well understood. One can learn
more about this problem, if the mismatching between the initial
and final QRPA (RQRPA) Hamiltonians is studied \cite{sims97}. 
It has been found that pn-RQRPA Hamiltonians demonstrate a better 
mutual agreement than the pn-QRPA ones \cite{sims97}. 

Recently, the validity of the RQRPA have been discussed within  schematic 
models, which are trying to simulate the realistic cases
either by analytical solutions or by a minimal computational
effort \cite{hirm96,eng97,del97,sam97}. 
Again it was  confirmed  that the renormalized QRPA is superior 
to the QRPA. But differences were found
between the exact and the RQRPA solutions after the point of collapse of 
 QRPA. However, it is not clear whether these results would  also hold 
for realistic calculations with  large model spaces, realistic
effective NN-interactions and by consideration of the two-vacua problem. 
In fact we do not know of any exactly solvable realistic model. 

We note that there is a principal difference between the $2\nu\beta\beta$-decay
and $0\nu\beta\beta$-decay modes from the nuclear physics point of view. 
In the $2\nu\beta\beta$-decay the correlation of 
two single nucleon beta decays in the nucleus are taken into account by
the interaction part of the nuclear Hamiltonian. For a correct treatment
of these correlations higher order terms in the boson (renormalized boson)
expansion of the nuclear Hamiltonian are important \cite{sims97}. 
Several procedures have been proposed which
outline the importance of the higher order effects 
\cite{rad91,stoi93,rad96,sam97}. It is expected that the consideration
of these higher order approximations will improve the correspondence
between the initial and final nuclear Hamiltonian and more reliable 
results can be obtained. 

The $0\nu\beta\beta$-decay is not a higher order effect in the boson
expansion of the nuclear Hamiltonian as the dominant correlation of the
two beta decay vertices in the nucleus comes from the exchange of lepton 
number violating particles. Therefore the renormalized QBA scheme should be
sufficient.  
We believe that the renormalized-QRPA is 
at present the most reliable method to deduce
the desired interesting lepton number non-conserving parameters from
the experimental lower limits on the half-lifes of 
$0\nu\beta\beta$-decay of heavier nuclei.


\subsection{Limits on lepton number violating parameters}

It is a common practice to impose limits on different lepton number
non-conserving parameters to assume that only one mechanism 
dominates at a time.  A more accurate procedure requires the construction of 
multi-dimensional plots. This is worth doing only when $0\nu\beta\beta$-decay
has been found in the future.  

In order to extract the limits on the effective neutrino mass $<m_\nu >$ 
and effective Majoron coupling $<g>$
from the experimental limits of $0\nu\beta\beta$-decay lifes the
nuclear matrix elements $M^{0\nu}_{GT}$ and $\chi_F$ have to be calculated.
We have evaluated these matrix elements for the  experimentally most 
interesting  A=76, 82, 96, 100, 116, 128, 130, 136 and 150 nuclear systems 
within the framework of  pn-RQRPA \cite{toi95,simn96}. 
The same model space and nuclear structure parameters have been used 
as those considered in \cite{fae98}. 
The calculated nuclear matrix elements for $g_{pp}=1.0 $ together 
with upper bounds on  $<m_\nu >$ and $<g>$ 
are presented in Table \ref{tabres.1}. We see that nowadays
$^{76}Ge$ provides 
the most stringent limit on  the effective neutrino mass: $<m_\nu > < 0.54 eV$.
This value is about twice smaller as the value obtained within the 
full-RQRPA \cite{simf97}, which include proton-neutron pairing  
correlations. The most stringent upper constraint on $<g>$ 
is associated with A=128 and is equal to $5.5\times 10^{-5}$.

As we mentioned already many different nuclear models 
(shell model \cite{ret95,hax84}, general seniority scheme \cite{eng89},
pn-QRPA \cite{eng88,mut89,tom91,pan96}, full-QRPA \cite{pan96}
and pn-, full-RQRPA \cite{simf97} and others ) have been used
for the calculation of the $0\nu\beta\beta$-decay matrix elements.
 Unfortunately, there is no possibility to assign an 
uncertainty to the calculated theoretical values. 
One can only claim that the renormalized QRPA predicts more reliable 
results in respect to the QRPA because of reasons discussed before.
A comparison between  the shell model 
and the QRPA (RQRPA) is not possible as both models work with 
fully different model spaces and nucleon-nucleon interactions. 
We remind that the QRPA is loosing sense
for a small model space and shell model studies 
for heavier open-shell nuclei are not possible. Thus 
one can expect that for the  A=48 system the shell model results are
more reliable and for other nuclear systems 
the RQRPA perhaps have  more predictive power. 

First, we discuss here the light Majorana neutrino exchange mechanisms. 
The $0\nu\beta\beta$-decay half-lifes for several nuclei 
calculated from some representative nuclear matrix elements of different
nuclear models are listed in Table \ref{tabres.2}. We have assumed 
that only one mechanism at a time is important and that
$<m_\nu > = 1 eV$,  $<\lambda > = 10^{-6}$, $<\eta > = 10^{-8}$ and
$<g> = 10^{-5}$. 
By glancing at  Table \ref{tabres.2} we see that most  theoretical works
have concentrated  on  A=76, 82, 128 and 130  nuclear systems
and that the calculated half-lifes agree within a factor  ten. 
It is remarkable that there is a large  disagreement between the two different
shell model calculations for $^{76}Ge$ and $^{82}Se$ \cite{hax84,ret95}. 
We note that from the theoretical point of view the best candidates 
for the experimental measurements are $^{150}Nd$ and $^{130}Te$ nuclei 
with the shortest half-lifes. The upper bound on
$<m_\nu >$,  $<\lambda >$, $<\eta >$ and $<g>$ can be deduced 
from the experimental limit on the $0\nu\beta\beta$-decay 
$T^{exp}_{1/2}$ given in Table \ref{tabint.1} with help of the
theoretical half-life $T^{theor}_{1/2}$ [see Table \ref{tabres.2}]
as follows:
\begin{equation}
 < I >~~ = ~~C_I ~~\sqrt{\frac{T^{theor}_{1/2}}{T^{exp}_{1/2}}},
\end{equation}
where $C_I$ takes the value 1 eV, $10^{-6}$, $10^{-8}$ and $10^{-5}$
for I = $m_\nu$, $\lambda $, $\eta $ and $g$, respectively. Thus, the most
restrictive  limits are as follows:
\begin{eqnarray}
<m_\nu > ~< ~~ 0.4 - 1.3 eV ~~[^{76}Ge,~ ref.~ 21], ~~~~~~~~~~
              0.7 - 1.5 eV ~~[^{128}Te, ~ ref.~ 35]\\
<\lambda >~ < ~~ (0.8-2.1)\times 10^{-6}~~[^{76}Ge,~ ref.~ 21] ~~~
                (2.5-7.3)\times 10^{-6}~~[^{136}Xe,~ ref.~ 37]\\
<\eta >~ <  ~~(0.4-1.8)\times 10^{-8}~~[^{76}Ge, ~ref.~ 21], ~~~
              (0.9-3.7)\times 10^{-8}~~[^{128}Te, ~ref.~ 35]\\
<g> ~ <  ~~(2.9-5.8)\times 10^{-5}~[^{128}Te,~ref.~ 35], ~~
         (9.8-22.)\times 10^{-5}~~[^{100}Mo,~ref.~ 29]
\label{vio.1}
\end{eqnarray}
The sensitivity  of different experiments
to these $<m_\nu >$,  $<\lambda >$, $<\eta >$ and
$<g> $. is drawn in Fig. \ref{fighis.1}. Currently, the Heidelberg-Moscow
experiment \cite{gue97}offers the most stringent limit for 
$<m_\nu >$,  $<\lambda >$, $<\eta >$ and the $^{128}Te$  experiment 
\cite{bern92} for $<g>$.

Second, we present nuclear matrix elements 
associated with the $R_p \hspace{-1em}/\;\:$  SUSY and heavy 
neutrino exchange mechanisms. 

The nuclear matrix 
elements governing the two-nucleon, one pion-exchange and two-pion 
exchange $R_p \hspace{-1em}/\;\:$  SUSY 
contributions to the $0\nu\beta\beta$-decay have been calculated 
only within the pn-QRPA, pn-RQRPA and the full-RQRPA approaches 
\cite{hkk96,fae97,fae98}. 
It has been found that the pion-exchange 
contribution dominates over the conventional two-nucleon one
and its contribution can be safely neglected \cite{fae97}. 
The calculated pion-exchange nuclear matrix elements ${\cal M}^{\pi N}$
for the $0\nu\beta\beta$-decay of various isotopes  within the
pn-RQRPA ($g_{pp}=1.0$) are presented in Table \ref{tabres.3} \cite{fae98}.
The two largest are  -1073 and -756 for A=150 and 100, respectively.
It has been found that the variation of ${\cal M}^{\pi N}$ on $g_{pp}$
do not exceed $15\%$ within the physical region of $g_{pp}$.
In the evaluation of ${\cal M}^{\pi N}$ 
the pion vertex structure coefficient was considered to be
$\alpha^{2\pi}=0.2$ (VIA) \cite{fae97}. 
We note that the value of ${\cal M}^{\pi N}$  calculated for $^{76}Ge$
within the full-RQRPA \cite{fae97} is suppressed by 
about  $30\%$ in comparison with pn-RQRPA. 
The upper bounds on $\eta^{exp}_{_{SUSY}}$  [see Eq. (\ref{limit.susy})]
have been deduced
by using the most stringent lower limits of $T^{exp}_{1/2}$ in 
Table \ref{tabint.1}. By glancing at Table \ref{tabres.3} we see that
$^{76}Ge$ provides the most stringent limit: 
$\eta^{exp}_{susy}$ $= 5.5\times 10^{-9}$. 
By using the Eqs. (\ref{susy.9}) and (\ref{susy.10}) one obtains for the
1st generation  $R_p \hspace{-1em}/\;\:$ Yukawa coupling constant 
\cite{fae98}
\begin{eqnarray}
\lambda'_{111} \le 1.3 \times 10^{-4}
\Big({m_{\tilde q}\over{100 ~{\text{GeV}}}} \Big)^2
 \Big({m_{\tilde g}\over{100 ~{\text{GeV}}}} \Big)^{1/2}, \\
\lambda'_{111} \le 9.3 \times 10^{-4}
\Big({m_{\tilde e}\over{100 ~{\text{GeV}}}} \Big)^2
 \Big({m_{\chi}\over{100 ~{\text{GeV}}}} \Big)^{1/2}.
\label{vio.2}
\end{eqnarray} 
These limits are much stronger than those previously known
and lie beyond the reach of the near future  accelerator experiments.
The size of $\lambda'_{111}$ depends on the masses
of the SUSY-partners. 
If the values of these masses would be around
their present experimental lower
limits $\sim 100$GeV \cite{par96}, one could constrain
the coupling to $\lambda'_{111}\leq 1.3 \cdot 10^{-4}$.
A conservative bound can
be set by assuming all the SUSY-masses being at the "SUSY-naturalness"
bound of 1~TeV, leading to $\lambda'_{111}\leq 0.04$.

The nuclear matrix element ${\cal M}_N$ associated with the exchange 
of heavy Majorana neutrino are shown for different
nuclear systems in Table \ref{tabres.3}.  They have been evaluated 
within the pn-RQRPA for the same nuclear model parameters 
as $M^{0\nu}_{GT}$ and $\chi_F$ in Table \ref{tabres.1}.
The two largest elements are  
-206 and -151 for for A=150 and 100, respectively. It is worthwhile to
note that these values can be reduced, if higher order terms of the
nucleon current are considered \cite{lyu92}. For the A=76 system
the most stringent limit on the relevant lepton number non-conserving
parameter has been found: $\eta_N < 2.6\times 10^{-8}$.


\section{Conclusion and Outlook}

In this review we have discussed  the recent progress in the field of 
the nuclear double beta decay and its present status. 

The $2\nu\beta\beta$-decay remains at the forefront of nuclear physics.
The established $2\nu\beta\beta$-decay half-lifes 
for a couple of isotopes constrain  nuclear theory and stimulate
its further development. 
There are still a number of open questions concerning the solution of the
nuclear many-body problem.
In order to increase the predictive power
of the nuclear many-body models it seems to be necessary to include better 
the Pauli exclusion 
principle in the evaluation of the $2\nu\beta\beta$-decay 
matrix elements. This is done in the  renormalized QRPA. It is an improvement
over the commonly
used QRPA approach. An improved description of the 
complete set of the intermediate nuclear states, which are  needed for the 
calculation of double beta decay, can  possibly also be obtained with
 growing computer power for example in the shell model approach. 

Additional experimental information about the $2\nu\beta\beta$-decay
and related processes is of great interest. We note that 
the NEMO experiment allows to perform a precise
measurement also of the energy and angular distributions of the outgoing 
electrons. Further, there is a chance to detect the
$2\nu\beta\beta$-decay to the excited $0^+$ and $2^+$ states of the 
final nucleus at the level of $10^{21}$-$10^{22}$ years with present 
low-background detectors. New $\beta\beta$ experiments are planned by the 
detection of the 
capture of two bound atomic electrons \cite{ste98}. A recent
theoretical analysis has shown that the experimental measurement of
the double beta decay induced in a neutrino beam can  be feasible due to an 
enhancement  due to a resonance in  the
corresponding amplitude \cite{gap97}. Such an experiment is planned 
for the double beta decay at the Kurchatow 
Institute \cite{gap97}. These future experiments are important 
because they complement the known informations from the
conventional $2\nu\beta\beta$-decay.

The $0\nu\beta\beta$-decay is a powerful tool for the study of
lepton number conservation in general. In particular,
it contributes to the search for the Majorana neutrino mass, the heavy 
W-boson (mainly responsible for a possible right-handed weak interaction), the
$R_p \hspace{-1em}/\;\:$ SUSY and leptoquarks  on the sensitivity 
level competitive to Large Hadron Colliders.
We remind that the study of rare decays at low energy is complementary 
to high energy physics experiments.

Searches for double beta decay are pursued actively for 
different nuclear isotopes, but the
$0\nu\beta\beta$-decay has been not yet observed, experimentally. 
The currently laboratory upper limits on the half-life
provide severe constraints on the effective 
Majorana neutrino mass $<m_\nu >$, on parameters of the left-right
symmetric models $<\lambda >$ and $<\eta >$, on the effective Majoron coupling
constant $<g>$, on the $R$-parity violating coupling constant
$\lambda'_{111}$ and on the the inverse effective mass of heavy neutrinos
${\eta_N}$ as follows:
\begin{eqnarray}
<m_\nu > ~ &<& ~~ 0.4 - 1.3 eV, 
~~~~~~~~~~~~~~~~~~~~~~<g> ~ <  ~~(2.9-5.8)\times 10^{-5},
\nonumber \\
<\lambda >~ &<& ~~ (0.8-2.1)\times 10^{-6},~~~~~~~~~~~~~~~
<\eta >~ <  ~~(0.4-1.8)\times 10^{-8}
\nonumber \\
\lambda'_{111}~~~ &\le & ~~1.3 \times 10^{-4}
\Big({m_{\tilde q}\over{100 ~{\text{GeV}}}} \Big)^2
 \Big({m_{\tilde g}\over{100 ~{\text{GeV}}}} \Big)^{1/2}, ~~~~
\eta_N ~<  ~2.6\times 10^{-8}
\label{out.1}
\end{eqnarray}
The uncertainty of these parameters 
is due to the ambiguity of nuclear matrix elements.

The above limits constitute our most stringent test of the 
lepton number conservation. 
It is worthwhile to notice that the $0\nu\beta\beta$-decay
imposes very restrictive bounds on the lepton number 
violating sector of R-parity violating SUSY models. The upper limit on 
$\lambda'_{111}$ (deduced from the pion-exchange
$R_p \hspace{-1em}/\;\:$ SUSY mode for the
$0\nu\beta\beta$-decay \cite{fae97}),
is significantly stronger than those previously known or expected
from the ongoing accelerator experiments.
The $0\nu\beta\beta$-decay constraints 
on lepton number violating parameters in Eq. (\ref{out.1}) must
be taken into account by theoreticians, when they build new 
theories of  Grand Unification.

In the $0\nu\beta\beta$-decay it is still possible to improve the 
existing limits. The extension of the ongoing  
experiments: ELEGANT ($^{48}Ca$) \cite{kum95},
Heidelberg-Moscow  ($^{76}Ge$) \cite{bau97},
NEMO ($^{100}Mo$, $^{116}Cd$) \cite{das95,arn96}, 
 the $^{130}Te$ cryogenic experiment \cite{ale94} and  
Caltech Neuchatel TPC $^{136}Xe$ \cite{bus96} is expected to reach 
the level of 0.1 eV. The prospect for exploration of SUSY in the next 
generation of the $0\nu\beta\beta$-decay experiments have been discussed in 
\cite{bedn97}. Recently, a new project GENIUS  for the measurement of the
$0\nu\beta\beta$-decay of $^{76}Ge$ have been proposed, which suppose 
to probe neutrino masses down to
$10^{-2}$-$10^{-3}$ eV by using 1 ton of enriched $^{76}Ge$ \cite{hell97}.
If the $0\nu\beta\beta$-decay will be discovered, it would be a major 
achievement, which would initiate new experimental activities both for rare
decays at 
low-energies and also at high energy accelerator facilities. In 
the case that the
lepton number violation is out of reach of near future $0\nu\beta\beta$-decay 
experiments, the improved upper limits will yield 
more stringent constraints for the Grand Unified Theories and Super Symmetric
(SUSY) models.


\newpage

\begin{table}[h]
\caption{Experimental results for $2\nu\beta\beta$-, 
$0\nu\beta\beta$- and $0\nu\beta\beta\phi$-decay modes
for A=48, 76, 82, 96, 100, 116, 128, 130, 136, 150, 238 and 242 nuclear
systems. gch.(rch.) - geochemical (radiochemical) data.}
\begin{tabular}{cccc}
nuclear &  
$T_{1/2}^{2\nu\beta\beta}(0^+_{g.s.}\rightarrow 0^+_{g.s.}$) &
$T_{1/2}^{0\nu\beta\beta}(0^+_{g.s.}\rightarrow 0^+_{g.s.}$) &
$T_{1/2}^{0\nu\beta\beta \Phi}(0^+_{g.s.}\rightarrow 0^+_{g.s.}$) \\
transition & [$y^{-1}$]  ref. & 
[$y^{-1}$] C.L. ref. & [$y^{-1}$] C.L. ref.  \\ \hline
$^{48}Ca\rightarrow{^{48}Ti}$ & 
$ > 3.6\times 10^{19}$ \cite{bard70} &  
$ > 2.0\times 10^{21}$ 80\% \cite{bard70}  & 
$ > 7.2\times 10^{20}$ 90\% \cite{bard70}  \\
 & $(4.3^{+2.4}_{-1.1}\pm 1.4)\times 10^{19}$ \cite{bal96} & 
 $ > 9.5\times 10^{21}$ $76\%$ \cite{you91} &  \\
$^{76}Ge\rightarrow{^{76}Se}$ 
 & $(1.77^{+0.01~~+0.13}_{-0.01~~-0.11})\times 10^{21}$ \cite{gue97} 
 & $> 1.1 \times 10^{25}$ $90\%$ \cite{bau97}
 & $> 7.91 \times 10^{21}$ \cite{gue97} \\
 & $(9.2^{+0.7}_{-0.4})\times 10^{20}$ \cite{avi91} & & \\ 
$^{82}Se\rightarrow{^{82}Kr}$ 
 & $(1.08^{+0.26}_{-0.06}) \times 10^{20}$ \cite{ell92}
 & $> 2.7 \times 10^{22}$ $68\%$ \cite{ell92}
 & $> 1.6 \times 10^{21}$ $68\%$ \cite{moe94} \\
 & $(1.2\pm 0.1)\times 10^{20}$  gch. \cite{lin88} & & \\
$^{96}Zr\rightarrow{^{96}Mo}$ 
 & $> 3.9 \times 10^{19}$ gch. \cite{kaw93}
 & $> 3.9 \times 10^{19}$ gch. \cite{kaw93}
 & $> 3.9 \times 10^{19}$ gch. \cite{kaw93} \\
$^{100}Mo\rightarrow{^{100}Ru}$ 
 & $(6.82^{+0.38}_{-0.53}\pm 0.68)\times 10^{18}$ \cite{sil97}
 & $> 1.23 \times 10^{21}$ $90\%$ \cite{sil97} 
 & $> 3.31 \times 10^{20}$ $90\%$ \cite{sil97} \\
 & $(9.5\pm 0.4\pm 0.9)\times 10^{18}$ \cite{das95} 
 & $> 6.4 \times 10^{21}$ $90\%$ \cite{das95} 
 & $> 5 \times 10^{20}$ $90\%$ \cite{das95} \\
 & $ 11.5^{+3.0}_{-2.0}$ $\times 10^{18}$ \cite{eji91}
 & $> 5.2 \times 10^{22}$ $68\%$ \cite{eji96}
 & $> 5.4 \times 10^{21}$ $68\%$ \cite{eji96} \\
 & $7.6^{+2.2}_{-1.4}$ $\times 10^{18}$ \cite{als97}
 & $> 2.2 \times 10^{22}$ $68\%$ \cite{als97} & \\
 & $ 3.3^{2.0}_{-1.0}$ $\times 10^{18}$ \cite{vas90}
 & $> 0.71 \times 10^{21}$ $68\%$ \cite{vas90} & \\
$^{116}Cd\rightarrow{^{116}Sn}$ 
 & $(2.7^{+0.5}_{-0.4}\pm 0.9)\times 10^{19}$ \cite{dane95}
 & $> 2.9 \times 10^{22}$ $90\%$ \cite{dane95} & \\
 & $(2.6 ^{+0.9}_{-0.5}\pm 0.35)\times 10^{19}$ \cite{kume94} 
 & $> 5.4 \times 10^{21}$ $68\%$ \cite{kume94} & \\
 & $(3.6\pm 0.35\pm 0.21)\times 10^{19}$ \cite{arn96} 
 & $> 5.0 \times 10^{21}$ $90\%$ \cite{arn96}
 & $> 1.2 \times 10^{21}$ $90\%$ \cite{arn96}
\\
%
$^{128}Te\rightarrow{^{128}Xe}$ 
 & $(7.7\pm 0.4)\times 10^{24}$ gch. \cite{bern92}
 & $> 7.7\times 10^{24}$ gch. \cite{bern92}
 & $> 7.7\times 10^{24}$ gch. \cite{bern92} \\
$^{130}Te\rightarrow{^{130}Xe}$ 
 & $(2.7\pm 0.1)\times 10^{21}$ gch. \cite{bern92}
 & $> 8.2 \times 10^{21}$ \cite{ale94}
 & $> 2.7 \times 10^{21}$ gch. \cite{bern92} \\
$^{136}Xe\rightarrow{^{136}Ba}$ 
 & $> 5.5 \times 10^{20}$ \cite{bus96} 
 & $> 4.2 \times 10^{23}$ \cite{bus96} 
 & $> 1.4 \times 10^{22}$ \cite{bus96}  \\ 
$^{150}Nd\rightarrow{^{150}Sm}$  
 & $(6.75^{+0.37}_{-0.42}\pm 0.68)\times 10^{18}$ \cite{sil97}
 & $> 1.22 \times 10^{21}$ $90\%$ \cite{sil97}
 & $> 2.82 \times 10^{20}$ $90\%$ \cite{sil97} \\
 & $(18.8^{+6.6}_{-3.9}\pm 1.9)\times 10^{18}$ \cite{art95}
 & $> 2.1 \times 10^{20}$ $ 90\%$ \cite{art95}
 & $> 1.7 \times 10^{20}$ $ 90\%$ \cite{art95} \\
 & $> 11 \times 10^{18}$ $90\%$ \cite{vas93} & & \\
$^{238}U\rightarrow{^{238}Pu}$  
 & $ (2.0\pm 0.6)\times 10^{21}$ rch. \cite{tur91}
 &  $ > 2.0 \times 10^{21}$ rch. \cite{tur91}
 &  $ > 2.0 \times 10^{21}$ rch. \cite{tur91} \\
$^{242}Pu\rightarrow{^{242}Cm}$  
 & $> 1.1 \times 10^{18}$ rch. \cite{moo92}
 & $> 1.1 \times 10^{18}$ rch. \cite{moo92}
 & $> 1.1 \times 10^{18}$ rch. \cite{moo92} \\
\end{tabular}
\label{tabint.1}
\end{table}

\begin{table}[t]
\caption{The transition operators  
${\cal O}_{I}({\mathbf{r}}_i,{\mathbf{r}}_j,
{\mathbf{\sigma}}_i,{\mathbf{\sigma}}_j)$ = $const \cdot$
$\frac{2}{\pi}\int_0^\infty {\cal P}_I(q,r) dq$
$ \cdot {\cal S}_I$(${\mathbf{\hat{r}}}_{ij}$, ${\mathbf{\hat{r}}}_{+ij}$,
${\mathbf{\sigma}}_i$, ${\mathbf{\sigma}}_j)$ (see Eqs. (\ref{neu.9}))
for the nuclear
matrix elements associated with exchange of light and heavy particles
are listed ( ${\mathbf{{r}}}_{ij}={\mathbf{r}}_i$ - ${\mathbf{r}}_j$, 
${\mathbf{{r}}}_{+ij}={\mathbf{r}}_i$ + ${\mathbf{r}}_j$,
${\mathbf{\hat{r}}}_{ij}={\mathbf{{r}}}_{ij}$$/|{\mathbf{{r}}}_{ij}|$,
${\mathbf{\hat{r}}}_{+ij}={\mathbf{{r}}}_{+ij}$$/|{\mathbf{{r}}}_{+ij}|$,
$r=|{\mathbf{\hat{r}}}_{ij}|$ ). The following notations are used:
R is nuclear radius. $j_i$ (i=1,2) are spherical Bessel functions.
$f(q^2)=1/(1+q^2/m_A^2)^2$ is the dipole 
nucleon form factor with cut-off $m_A$=850 MeV.
$m_\pi$ and $m_p$ are the pion and  
nucleon masses, respectively. $\overline{A}$ 
is the closure energy, which can be found for different nuclei
e.g. in [79]. ${\mathbf{S}}_{ij}=$$3 {\mathbf{\sigma}}_i$ $\cdot$ 
${\mathbf{\hat{r}}}_{ij} ~$
$ {\mathbf{\sigma}}_j$ $\cdot$ ${\mathbf{\hat{r}}}_{ij}$ -
 ${\mathbf{\sigma}}_i$ $\cdot$ ${\mathbf{\sigma}}_j$  
is the tensor operator. ${\mathbf{\sigma}}_{-ij}=$
${\mathbf{\sigma}}_i-{\mathbf{\sigma}}_j$.
$g_V = 1.0$,  $ g_A=1.24$ and $\mu_\beta=4.7$}
\label{tabneu.1}
\begin{tabular}{lccc}
 M.E. & const & ${\cal P}(q,r)~$ & $ {\cal S}$
(${\mathbf{\hat{r}}}_{ij}$, ${\mathbf{\hat{r}}}_{+ij}$,
${\mathbf{\sigma}}_i$, ${\mathbf{\sigma}}_j)$ \\ \hline
 & \multicolumn{3}{c}{ light particle exchange mechanism}\\
$M^{0\nu}_{GT}$ & R & 
$  j_0(q r) \frac{q}{q+\overline{A}} f^2(q^2) $  & 
${\mathbf{\sigma}}_i \cdot {\mathbf{\sigma}}_j$ \\
$\chi_F$ & $\big(\frac{g_V}{g_A}\big)^2 \frac{1}{M^{0\nu}_{GT}}$ $R$  & 
$   j_0(q r) \frac{q}{q+\overline{A}} f^2(q^2) $  & 1  \\
$\chi_{GT'}$ &  $\frac{1}{M^{0\nu}_{GT}}$ $R^2$ & 
$ j_1(q r) \frac{q^2}{q+\overline{A}} f^2(q^2) $  
& ${\mathbf{\sigma}}_i \cdot {\mathbf{\sigma}}_j$ \\
$\chi_{F'}$ &  
$\big(\frac{g_V}{g_A}\big)^2\frac{1}{M^{0\nu}_{GT}}$ $R^2$ &
$ j_1(q r) \frac{q^2}{q+\overline{A}} f^2(q^2) $  
 & 1 \\
$\chi_{T'}$ &  $\frac{1}{M^{0\nu}_{GT}}$ $\frac{R^2}{3}$ & 
$  j_1(q r) \frac{q^2}{q+\overline{A}} f^2(q^2) $  & 
${\mathbf{S}}_{ij}$ \\
$\chi_{GT\omega}$ &  $\frac{1}{M^{0\nu}_{GT}}$ $R$ & 
$  j_0(q r) \frac{q^2}{(q+\overline{A})^2} f^2(q^2) $   
& ${\mathbf{\sigma}}_i \cdot {\mathbf{\sigma}}_j$ \\
$\chi_{F\omega}$ & $\big(\frac{g_V}{g_A}\big)^2\frac{1}{M^{0\nu}_{GT}}$ $R$ &
$ j_0(q r) \frac{q^2}{(q+\overline{A})^2} f^2(q^2) $   
 &  1 \\
$\chi_{P}$ &  $\big(\frac{g_V}{g_A}\big)$ $\frac{1}{M^{0\nu}_{GT}}$ $R^2$ & 
$ j_1(q r) \frac{q^2}{q+\overline{A}} f^2(q^2) $  
& $i {\mathbf{\sigma}}_{-ij}\cdot \big({\mathbf{\hat{r}}}_{ij} 
\times \frac{{\mathbf{\hat{r}}}_{+ij}}{R}\big)$ \\
$\chi_{R}$ &  $\frac{\mu_\beta}{3} \big(\frac{g_V}{g_A}\big)$ 
$\frac{1}{M^{0\nu}_{GT}}$ $\frac{R^2}{m_p}$ & 
$ (j_0(qr)-\frac{2j_1(q r)}{qr}) \frac{q^3}{q+\overline{A}} f^2(q^2) $  
 & ${\mathbf{\sigma}}_i \cdot {\mathbf{\sigma}}_j$ \\
 & \multicolumn{3}{c}{heavy particle exchange mechanism}\\
 ${\cal M}_{GT N}$ & $\frac{R}{m^2_A}$ & 
$ j_0(qr) q^2 f^2(q^2) $ & 
${\mathbf{\sigma}}_i \cdot {\mathbf{\sigma}}_j$ \\
 ${\cal M}_{F N}$ & $\frac{R}{m^2_A}$ & 
$ j_0(qr) q^2 f^2(q^2) $ & 1 \\
 ${\cal M}_{GT'}$ & $\frac{R}{m^4_A}$ & 
$ j_0(qr) q^4 f^2(q^2) $ &  
${\mathbf{\sigma}}_i \cdot {\mathbf{\sigma}}_j$ \\
 ${\cal M}_{F'}$ & $\frac{R}{m^4_A}$ & 
$ j_0(qr) q^4 f^2(q^2) $ &  
1 \\
 ${\cal M}_{T'}$ & $\frac{R}{m^4_A}$  & 
$ j_2(qr) q^4 f^2(q^2) $ &  
${\mathbf{S}}_{ij}$ \\
 ${\cal M}_{GT-1\pi}$ & $-\frac{R}{m^2_\pi}$ & 
$ j_0 (qr) \frac{q^4}{(q^2+m^2_\pi)} f^2(q^2) $ & 
${\mathbf{\sigma}}_i \cdot {\mathbf{\sigma}}_j$ \\
 ${\cal M}_{T-1\pi}$ & $\frac{R}{m^2_\pi}$ & 
$ j_2 (qr) \frac{q^4}{(q^2+m^2_\pi)} f^2(q^2) $ & 
${\mathbf{S}}_{ij}$ \\
 ${\cal M}_{GT-2\pi}$ & $- 2  R$ & 
$ j_0 (qr) \frac{q^4}{(q^2+m^2_\pi)^2} f^2(q^2) $ & 
${\mathbf{\sigma}}_i \cdot {\mathbf{\sigma}}_j$ \\
 ${\cal M}_{T-2\pi}$ & $ 2 R$ &  
$ j_2 (qr) \frac{q^4}{(q^2+m^2_\pi)^2} f^2(q^2)  $ & 
${\mathbf{S}}_{ij}$ \\ 
\end{tabular}
\end{table}

\begin{table}[t]
\caption{The nucleon structure coefficients of the two-nucleon mode
for the  $R_p \hspace{-1em}/\;\:$  SUSY mechanism for the
 $0\nu\beta\beta$-decay
calculated within the non-relativistic quark  (QM) and 
Bag models.  They have been calculated using 
Table I and Eqs. (54-57) of ref. [58].}
\label{tabsus.1}
\begin{tabular}{lccccccc}
Model & i & \hspace{0.3cm}$a^{(0)}_{V-i}$ \hspace{0.3cm}
& \hspace{0.3cm} $a^{(0)}_{A-i}$ \hspace{0.3cm} & 
\hspace{0.3cm} $a^{(1)}_{V-i}$ \hspace{0.3cm} &  
\hspace{0.3cm} $a^{(1)}_{A-i}$ \hspace{0.3cm} &  
\hspace{0.3cm} $a^{}_{T-i}$ \hspace{0.3cm} \\ \hline
Non-rel. & $\tilde{q}$ & 0.145 & -1.198 & 8.221 & 0.306 & 1.408 \\ 
 QM     & $\tilde{f}$ & 0.145 &  0.0   & 0.0   & -0.837 & 0.837 \\
 & & & & & & \\
 Bag    & $\tilde{q}$ & 0.242 & -1.322 & 2.511 & -0.670 & 1.061 \\
 model  & $\tilde{f}$ & 0.242 &  0.0   & 0.0   & -0.931 & 0.931 \\
\end{tabular}
\end{table}

\begin{table}[t]
\caption{The matrix elements $M^{0\nu}_{GT}$ and $\chi_F$ of the  
$0\nu\beta\beta$-decay for $^{48}{{Ca}}$,
$^{76}{{Ge}}$, $^{82}{{Se}}$, $^{96}{{Zr}}$, 
$^{100}{{Mo}}$, $^{116}{{Cd}}$, $^{128}{{Te}}$,
$^{130}{{Te}}$ and $^{136}{\text{Xe}}$ calculated in the
framework of pn-RQRPA. The upper limits on the lepton number non-conserving 
parameter $<m_\nu >$ ($<g>$) have been extracted from 
the best presently available experimental 
$0\nu\beta\beta$-decay ($0\nu\beta\beta\phi$-decay) 
half-lifes [see Table I ]. }
\label{tabres.1}
\begin{tabular}{cccccccccc}
\multicolumn{10}{c}{ Nucleus} \\\cline{2-10}
 & $^{76}Ge$ & $^{82}Se$ & $^{96}Zr$ & $^{100}Mo$ &
 $^{116}Cd$ & $^{128}Te$ & $^{130}Te$ & $^{136}Xe$ & $^{150}Nd$ \\ \hline 
$M^{0\nu}_{GT}$ & 2.80 & 2.66 & 1.54 & 3.30 & 2.08 & 2.21 & 1.84 & 0.70
 & 3.37 \\
$\chi_F$ & -0.29 & -0.28 & -0.29 & -0.25 & -0.24 & -0.34 & -0.36 & -0.46 
 & -0.34 \\
$<m_\nu >$ & 0.54 & 5.5 & 167. & 2.5 & 5.2 & 1.5 & 10.7 & 3.5 & 7.1 \\
$<g>\times 10^4$ & 4.0 & 3.3 & 22. & 1.1 & 25. & 0.55 & 3.0 & 3.2 & 1.8 \\
\end{tabular}
\end{table}

\begin{table}[t]
\caption{
Calculated $0\nu\beta\beta$-decay half-lifes for
 $ <m_\nu > = 1 eV$, $<\lambda > = 10^{-6}$, $<\eta > = 10^{-8}$ and
$<g> = 10^{-5}$ by assuming that one mechanism dominates at a time. 
The results have been obtained
within the  following nuclear models: shell model [1,156],
generalized seniority scheme [191], 
quark confinement model + pn-QRPA [132],
pn-QRPA [79,167,168] and [126]a,
full-QRPA [126]b,
 pn-RQRPA (p.w. - present work) and full-RQRPA [109].
}
\begin{tabular}{ccccccccccc} 
\multicolumn{11}{c}{ 
$T^{0\nu-theor.}_{1/2}(<m_\nu >, <\lambda >, <\eta >, <g>)$
[$years^{-1}$]} \\\cline{2-11}
 & $^{48}Ca$ & $^{76}Ge$ & $^{82}Se$ & $^{96}Zr$ & $^{100}Mo$ &
 $^{116}Cd$ & $^{128}Te$ & $^{130}Te$ & $^{136}Xe$ & $^{150}Nd$ \\ 
Ref. & $10^{24}$ & $10^{24}$ & $10^{24}$ & $10^{24}$ & 
   $10^{24}$ & $10^{24}$ & $10^{25}$ & $10^{24}$ & $10^{24}$ & 
 $10^{22}$ \\ \hline
\multicolumn{11}{c}{ $<m_\nu > = 1 eV ~~ <\lambda > = 0~~
<\eta > = 0 ~~ <g> = 0$} \\
\cite{ret95} & 6.42 & 17.4 & 2.40 & & & & & & 12.1 & \\
\cite{hax84} & 3.17 & 1.68 & 0.58 & & & & 0.40 & 0.16 & & \\
\cite{eng89} &  & 2.30 & 0.92 & & & & 0.45 & 0.24 & &  \\
\cite{eng88} &  & 14.0 & 5.60 & & & & 1.50 & 0.66 & 3.30 & \\
\cite{suh91} & & 4.06 & 1.43 & & & & 1.80 & 0.83 & & \\
\cite{mut89} & & 2.33 & 0.60 & & 1.27 & & 0.77 & 0.49 & 2.21 & 3.37 \\
\cite{tom91} & & 2.16 & 0.61 & & 0.26 & & 0.98 & 0.54 & 1.40 & 4.45 \\
 p.w. & & 3.15 & 0.80 & 1.09 & 0.34 & 0.77 & 1.63 & 0.94 & 5.29 & 6.11 \\
\cite{simf97} & & 8.95 & & & 0.25 & 0.70 & 1.09 & & 8.76 & \\
\multicolumn{11}{c}{ $<m_\nu > = 0~~ <\lambda > = 10^{-6}~~
<\eta > = 0 ~~ <g> = 0$} \\
\cite{ret95} & 7.45 & 50.2 & 3.25 & & & & & & 22.2 & \\
\cite{suh91} &  & 7.75 & 1.14 & & & & 14.8 & 0.89 &  & \\
\cite{mut89} &  & 7.35 & 0.99 & & 0.95 & & 13.5 & 0.95 & 4.90 & 3.73 \\
\cite{tom91} &  & 8.02 & 1.07 & & 0.55 & & 21.1 & 1.18 & 3.47 & 6.71 \\
\cite{pan96}a & 2.71 & 8.90 & 2.08 & 0.94 & 30.6 & 39.1 & 22.7 & 1.34 & 
2.73 &  \\
\cite{pan96}b & 27.9 & 41.2 & 4.39 & 27.7 & 10.3 & 10.8 & 165. & 2.22 & 
4.42 &  \\
\multicolumn{11}{c}{ $<m_\nu > = 0~~ <\lambda > = 0~~
<\eta > = 10^{-8} ~~ <g> = 0$} \\
\cite{ret95} & 6.42 & 27.2 & 6.24 & & & & & & 22.2 & \\
\cite{suh91} &  & 36.7 & 11.1 & & & & 10.7 & 5.92 &  & \\
\cite{mut89} &  & 2.25 & 0.65 & & 0.28 & & 0.67 & 0.44 & 1.21 & 3.39 \\
\cite{tom91} &  & 2.82 & 0.76 & & 0.34 & & 0.85 & 0.54 & 1.20 & 9.13 \\
\cite{pan96}a & 15.1 & 3.10 & 6.51 & 1.48 & 3.44 & 19.2 & 1.20 & 0.62 & 
1.23 &  \\
\cite{pan96}b & 43.2 & 22.8 & 5.16 & 7.95  & 102. & 83.2 & 1.90 & 1.05 & 
0.96 &  \\
\multicolumn{11}{c}{ $<m_\nu > = 0~~ <\lambda > = 0 ~~
<\eta > = 0 ~~ <g> = 10^{-5} $} \\
\cite{ret95} & 7.96 & 70.9 & 5.24 & & & & & & 32.5 & \\
\cite{hax84} & 3.93 & 6.85 & 1.27 & & & & 5.70 & 4.20 & & \\
\cite{suh91} & & 16.5 & 3.13 & & & & 25.7 & 2.19 & & \\
\cite{eng89} &  & 9.36 & 2.01 & & & & 6.41 & 0.63 & &  \\
\cite{eng88} &  & 57.0 & 12.2 & & & & 21.4 & 1.73 & 8.86 & \\
\cite{mut89} & & 9.49 & 1.32 & & 2.60 & & 11.1 & 1.28 & 5.94 & 5.30 \\
\cite{tom91} & & 8.77 & 1.32 & & 0.53 & & 13.9 & 1.42 & 3.76 & 6.99 \\
 p.w. & & 12.8 & 1.75 & 1.92 & 0.68 & 1.76 & 23.2 & 2.47 & 14.2 & 9.61 \\
\cite{simf97} & & 36.4 & & & 0.52 & 1.61 & 15.0 & & 23.5 & \\
\end{tabular}
\label{tabres.2}
\end{table}

\begin{table}[t]
\caption{The nuclear matrix elements for SUSY (${\cal M}^{\pi N}$) 
and heavy neutrino 
exchange (${\cal M}_N$) $0\nu\beta\beta$-decay (see Table \ref{tabneu.1}) 
for A=76,82,96,100, 116, 128, 130, 136 and 150 nuclear systems 
calculated within pn-RQRPA [153].
The lepton number non-conserving parameters 
$\eta_{_{SUSY}}$
and $\eta_{N}$ have been deduced
from the most stringent experimental limits of $0\nu\beta\beta$-decay 
lifes [see Table \ref{tabint.1}]. 
The upper bounds on SUSY coupling constant 
${\acute{\lambda}}_{111}$
have been obtained by using $ m_{\tilde{q}} = m_{\tilde{g}} =
m_{\tilde{\chi}}=100~ GeV $ [ see Eqs. (\ref{susy.9}) and (\ref{susy.10})].  
$T^{0\nu-susy}_{1/2}$ and 
$T^{0\nu-heavy}_{1/2}$ are the  calculated half lifes 
assuming  $\eta_{_{SUSY}}= 10^{-9}$ and $\eta_N = 10^{-8}$, respectively. }
\label{tabres.3}   
\begin{tabular}{cccccccccc}
\multicolumn{10}{c}{ Nucleus} \\\cline{2-10}
 & $^{76}Ge$ & $^{82}Se$ & $^{96}Zr$ & $^{100}Mo$ &
 $^{116}Cd$ & $^{128}Te$ & $^{130}Te$ & $^{136}Xe$ & $^{150}Nd$ \\ \hline
\multicolumn{10}{c}{ $R_p \hspace{-1em}/\;\:$  SUSY } \\ 
${\cal M}^{\pi N}\times 10^{-2}$ & -6.19 & -5.75 & -4.35 & -7.56 &
-4.33 & -6.98 & -6.35 & -3.72 & -10.7 \\
$T^{0\nu-susy}_{1/2}\times 10^{-25}$ & 33. & 8.6 & 7.1 & 3.0 &
 8.5 & 93. & 4.5 & 12. & 0.32 \\
$\eta_{susy} ~ \times 10^{8}$ & 0.55 & 5.6
 & 135. & 2.4 & 5.4 & 1.1 & 7.4 & 1.7 & 5.2 \\
${\acute{\lambda}}_{111} \times 10^{4}$ 
 & 0.79 & 2.5 & 12. & 1.7 &
 2.5 & 1.1 & 2.9 & 1.4 & 2.4 \\
\multicolumn{10}{c}{ heavy Majorana neutrino } \\
${\cal M}_N\times 10^{-2}$ & -1.30 & -1.20 & -0.84 & -1.51 & -0.91 &
-1.37 & -1.23 & -0.72 & -2.06  \\
$T^{0\nu-heavy}_{1/2}\times 10^{-25}$ & 7.4 & 2.0 & 1.9 & 0.76 &
1.9 & 24. & 1.2 & 3.2 & 0.088 \\  
$ \eta_N \times 10^{7}$ & 0.26 & 2.7 & 70. & 1.2 & 2.6 & 0.56 &
 3.8 & 0.88 & 2.70 \\ 
\end{tabular}
\end{table}

\newpage

\begin{figure}
\caption{Mechanisms for the $0\nu\beta\beta$-decay associated with 
the exchange of light Majorana neutrino: 
(a) the light neutrino mass mechanism, (b) 
the Majoron mechanism, (c-d) right-handed current mechanisms.
The following notation is used: $u_{L(R)}$, $d_{L(R)}$ and $e_{L(R)}$
are left- (right-) handed u-quark, d-quark and electron, respectively.
$W$ is vector boson (light or hypothetical heavy) and $\nu_i$ (i=1,2...)
is the Majorana neutrino). 
}
\label{figzero.1}
\end{figure}

\begin{figure}
\caption{
Feynman graphs for the supersymmetric contributions to 
$0\nu\beta\beta$-decay. $u_L$, $d_R$ and $e_L$ have the same meaning 
as in Fig. \ref{figzero.1}. 
${\tilde{u}}_L$, ${\tilde{d}}_R$ and ${\tilde{e}}_L$
are left-handed u-squark, right-handed d-squark and left-handed
selectron, respectively. $\chi$ and $\tilde{g}$ are neutralinos and
gauginos, respectively.
}
\label{figsus.1}
\end{figure}

\begin{figure}
\caption{\label{su_pro_cont}%
Two-nucleon (a), one-pion (b) and two-pion (c) mode of neutrinoless double
beta decay in $R$-parity violating SUSY extensions of the standard model.
}
\label{figsus.2}
\end{figure}

\begin{figure}[t]
\caption{The nuclear 
matrix element $M_{GT}$ for the $2\nu\beta\beta$-decay of $^{76}Ge$ 
calculated within the pn-QRPA (a) and pn-RQRPA (b)  is plotted as  a
function of the particle-particle coupling constant $g_{pp}$.
The dashed line corresponds to the 9-level model space
(the full $3-4\hbar\omega$ major oscillator shells),
the dot-dashed line to the 12-level model space
(the full $2-4\hbar\omega$ major oscillator shells)
and the solid line to the 21-level model space
(the full $0-5\hbar\omega$ major oscillator shells).} 
\label{res.1}
\end{figure}

\begin{figure}[htb]
\caption{The calculated nuclear matrix element 
$M^{0\nu}_{mass}=M^{0\nu}_{GT}(1-\chi_F)$
for the $0\nu\beta\beta$-decay of $^{76}Ge$,  $^{100}Mo$,  $^{116}Cd$,  
$^{128}Te$ and $^{136}Xe$  as a function of the
particle-particle interaction strength $g_{pp}$. In (a) 
$M^{0\nu}_{mass}$ has been calculated with the pn-QRPA method, in (b)
with the full-RQRPA . }
\label{res.2}
\end{figure}

\begin{figure}
\caption{The sensitivity of different experiments to the lepton-number
violating parameters 
$<m_\nu >$,  $<\lambda >$, $<\eta >$, $<g>$, $\eta_{_{SUSY}}$
and $\eta_{N}$. In construction of the histograms the best available 
limits on $0\nu\beta\beta $- and $0\nu\beta\beta\phi $-decay
half-lifes from Table I  have been used. The open and black bars in
(a-d) correspond to the smallest and largest  value of $T^{theor}_{1/2}$ 
for a given nucleus in Table V, respectively. 
}
\label{fighis.1}
\end{figure}


\begin{references}
\bibitem{hax84} Haxton W C and Stephenson G S \Journal{\PPNP}{12}{409}{1984}
\bibitem{gla61}  Glashow S L  \Journal{\NP}{22}{579}{1961}
\bibitem{wei67} Weinberg S \Journal{\PRL}{19}{1264}{1967}
\bibitem{sal68} Salam A {\em in Elementary Particle Theory:
 Relativistic Groups and Analyticity, Proceeding of the Eight
Nobel Symposium}, ed. by N. Svartholm (Almqvist and Wiksell, Stokholm),
1968 p. 367
\bibitem{may35} Goeppert-Meyer M, \Journal{\PR}{48}{512}{1935}
\bibitem{barb90} Barabash A S, \Journal{\JETL}{51}{207}{1990}
\bibitem{kopy90} Barabash A S, Kopylov A V and Cherovsky V I, 
      \Journal{\PLB}{249}{186}{1990}
\bibitem{blum92} Blum et al, \Journal{\PLB}{275}{506}{1992}
\bibitem{piep94} Piepke et al, \Journal{\NPA}{577}{493}{1994}
\bibitem{barb95} Barabash et al, \Journal{\PLB}{345}{408}{1995}
\bibitem{suc93} Suhonen J and Civitarese O, \Journal{\PLB}{308}{212}{1993}
\bibitem{suc94} Suhonen J and Civitarese O, \Journal{\PRC}{49}{3055}{1994}
\bibitem{sto94} Stoica S, \Journal{\PRC}{49}{2240}{1994}
\bibitem{dhi95} Dhiman S K and Raina P K, \Journal{\PRC}{50}{2660}{1995}
\bibitem{sch97} Schwieger J, \v Simkovic F, Faessler A and Kami\'nski W A,
   \Journal{\JPG}{23}{1647}{1997}
\bibitem{bara97} Suhonen J et al, \Journal{\ZPA}{358}{297}{1997}
\bibitem{bard70} Bardin R K, Gollon P J, Ullman J D and Wu C S, 
\Journal{\NPA}{158}{337}{1970}
\bibitem{bal96} Balysh A et al, \Journal{\PRL}{77}{5186}{1996}
\bibitem{you91} Ke You et al. \Journal{\PLB}{265}{53}{1991}
\bibitem{gue97} G\"unter et al, \Journal{\PRD}{55}{54}{1997}
\bibitem{bau97} Baudis L et al, \Journal{\PLB}{407}{219}{1997}
\bibitem{avi91} Avignone F T et al, \Journal{\PLB}{526}{559}{1991}
\bibitem{ell92} Elliot  S R et al, \Journal{\PRC}{46}{1535}{1992}
\bibitem{lin88} Lin W J et al, \Journal{\NPA}{481}{484}{1988} 
\bibitem{kaw93} Kawashima A, Takahashi K and Masuda A, 
  \Journal{\PRC}{47}{2452}{1993}
\bibitem{sil97} De Silva A, Moe M K, Nelson M A and Vient M A, 
  \Journal{\PRC}{56}{2451}{1997}
\bibitem{das95} Dassi\'e et al. (NEMO Collaboration), 
  \Journal{\PRD}{51}{2090}{1995}
\bibitem{eji91} Ejiri  H et al, \Journal{\JPG}{17}{155}{1991}
\bibitem{eji96} Ejiri  H et al, \Journal{\NPA}{611}{85}{1996}
\bibitem{als97} Alston-Garnjost M et al, \Journal{\PRC}{55}{474}{1997}
\bibitem{vas90} Vasil\'ev S I et al, \Journal{\JETL}{51}{622}{1990}
\bibitem{dane95} Danevich F A et al, \Journal{\PLB}{344}{72}{1995}
\bibitem{kume94} Kume K et al, \Journal{\NPA}{577}{405}{1994}
\bibitem{arn96} Arnold R et al, \Journal{\ZPC}{72}{239}{1996}
\bibitem{bern92} Bernatovicz T et al, \Journal{\PRL}{69}{2341}{1992};
  \Journal{\PRC}{47}{806}{1993}
\bibitem{ale94} Alessandrello A et al, \Journal{\NPBP}{35}{366}{1994} 
\bibitem{bus96} Busto J, \Journal{\NPBP}{48}{251}{1996}
\bibitem{art95} Artemiov V et al, \Journal{\PLB}{345}{564}{1995}
\bibitem{vas93} Vasil\'ev et al, \Journal{JETL}{58}{178}{1993}
\bibitem{tur91} Turkevich  A L, Economou T E and Gowan G A, 
  \Journal{\PRL}{67}{3211}{1991}
\bibitem{moo92} Moody K J, Lougheed R W, Hulet E K, 
  \Journal{\PRC}{46}{2624}{1992}
\bibitem{fur39} Fury W \Journal{\PR}{56}{1184}{1939}
\bibitem{moh91} Mohapatra R N and Pal P B {\em  Massive neutrinos in 
Physics and Astrophysics}, World Scientific, Singapore, 1991.
\bibitem{val92} Valle J W F {\em Workshop on Particles and Phenomena of
Fundamental Interactions} published in Jorge Swieca Summer School on 
Nuclear Physics 1995, hep-ph/9603307.
\bibitem{moh92} Mohapatra R N \Journal{\PRD}{46}{2990}{1992}
\bibitem{geo74} Georgi H M and Glasow S L \Journal{\PRL}{32}{438}{1974}
\bibitem{sch84} Schepkin M G \Journal{\SPU}{27}{555}{1984}
\bibitem{doi85} Doi M, Kotani T and Takasugi E \Journal{\PTPS}{83}{1}{1985}
\bibitem{ver86} Vergados J D \Journal{\PRP}{133}{1}{1986}
\bibitem{chi81} Chikashige T, Mohapatra R N and Peccei R D \Journal{\PRL}
  {45}{1926}{1980}; \Journal{\PLB}{98}{265}{1981}
\bibitem{gel81} Gelmini G B and Roncadelli M \Journal{\PLB}{99}{411}{1981}
\bibitem{geo81} Georgi H M, Glashow S L and Nussinov S 
   \Journal{\NPB}{193}{297}{1981}
\bibitem{sch82} Schechter J and Valle J W F \Journal{\PRD}{25}{774}{1982}
\bibitem{ber92} Berezhiani Z G, Smirnov A Yu and Valle J W F 
   \Journal{\PLB}{99}{291}{1992}
\bibitem{hir96} Hirsch M, Klapdor-Kleingrothaus H V, Kovalenko S G,
   P\"as H 
   
   \Journal{\PLB}{372}{8}{1996} 
\bibitem{moh86} Mohapatra R \Journal{\PRD}{34}{3457}{1986}
\bibitem{ver87} Vergados J D \Journal{\PLB}{184}{55}{1987}
\bibitem{hkk96} Hirsch M, Klapdor-Kleingrothaus H V and Kovalenko S G 
    \Journal{\PRD}{53}{1329}{1996}; \Journal{\PRL}{75}{17}{1995}
\bibitem{fae97} Faessler A, Kovalenko S, \v{S}imkovic F and Schwieger J,
    \Journal{\PRL}{78}{183}{1997}; Proceeding of the Int. Workshop on 
    Non-Accelerator New Physics (NANP'97), Dubna, Russia, June 1997, 
    1998 Phys. Atom. Nucl. {\bf 61} 1329 
\bibitem{hir97} Hirsch M, Klapdor-Kleingrothaus H V, Kovalenko S G,
    hep-ph/9707207
\bibitem{buc87} Buchm\"uller W, R\"uckl R and Wyler D, 
    \Journal{\PLB}{191}{442}{1987} 
\bibitem{lpq96} Hirsch M, Klapdor-Kleingrothaus H V, Kovalenko S G,
    \Journal{\PRD}{54}{4207}{1996}
\bibitem{vog86} Vogel P and Zirnbauer M R, \Journal{\PRL}{57}{3148}{1986}
\bibitem{civ87} Civitarese O, Faessler A and Tomoda T, 
     \Journal{\PLB}{194}{11}{1987}
\bibitem{mut88} Muto K and Klapdor H V, \Journal{\PLB}{201}{420}{1988};
     Muto K, Bender E and Klapdor H V, \Journal{\ZPA}{334}{177}{1989}
\bibitem{cheo93} Cheoun M K, Bobyk A, Faessler A, \v Simkovic F 
  and Teneva G, \Journal{\NPA}{561}{74}{1993}; \Journal{\NPA}{564}{329}{1993};
  Cheoun M K, Faessler A, \v Simkovic F, Teneva G and Bobyk A, 
  \Journal{\NPA}{587}{301}{1995}
\bibitem{rad91} Raduta A A, Faessler A, Stoica S and Kami\'nski W A, 
  \Journal{\PLB}{254}{7}{1991}; Raduta A A, Faessler A and Stoica,
  \Journal{\NPA}{534}{149}{1991}
\bibitem{civ91} Civitarese O, Faessler A, Suhonen J and Wu X R, 
  \Journal{\NPA}{524}{404}{1991}
\bibitem{krm93} Krmpoti\' c F, Mariano A, Kuo T T S and Nakayama K,
  \Journal{\PLB}{319}{393}{1993}
\bibitem{toi95} Toivanen J and Suhonen J, \Journal{\PRL}{75}{410}{1995};
  \Journal{\PRC}{55}{2314}{1997}
\bibitem{simn96} Schwieger J, \v Simkovic F and Faessler A, 
  \Journal{\NPA}{600}{179}{1996};
Schwieger J, Thesis Tuebingen 1997
\bibitem{krm96} Krmpoti\'c F,  Kuo T T S, Mariano A, de Passos E J V
  and de Toledo Piza A F R, \Journal{\NPA}{612}{223}{1997};
  \Journal{\FIZB}{5}{93}{1996}
\bibitem{hirm96} Hirsch J G, Hess P O and Civitarese O, 
  \Journal{\PRC}{54}{1976}{1996}; \Journal{\PRC}{56}{199}{1997};
  \Journal{\PLB}{390}{36}{1997}
\bibitem{simm97} \v Simkovic F and Pantis G, \Journal{\CJF}{48}{235}{1998}
\bibitem{sims97} \v Simkovic F, Pantis G and Faessler A, preprint 
  nucl-th/9711060 and 1998 {\em Yad. Fiz.} {\bf 61} (to be published); 
 \v Simkovic F, Pantis G and Faessler A, 1998
 {\em Prog. Part. Nucl. Phys.} 40 (to be published) 
\bibitem{pri81} Primakoff H and Rosen S P, \Journal{\ARNPS}{31}{145}{1981}
\bibitem{boe84} Boehm F and Vogel P, \Journal{\ARNPS}{34}{125}{1984}
\bibitem{avi88} Avignone F T III and Brodzinski R L, 
   \Journal{\PPNP}{21}{99}{1988}
\bibitem{tom91} Tomoda T, \Journal{\RPP}{54}{53}{1991}
\bibitem{gro90} Grotz K and Klapdor-Kleingrothaus H V, 
  {\it The Weak Interactions in Nuclear, Particle and Astrophysics}
  (Adam Hilger, Bristol, New York, 1990)
\bibitem{moe94} Moe M and Vogel P, \Journal{\ARNPS}{44}{247}{1994}


\bibitem{maj37} Majorana E, \Journal{\NCA}{14}{171}{1937}
\bibitem{pon57} Pontecorvo B, \Journal{\ZETF}{33}{549}{1957}
\bibitem{pon58} Pontecorvo B, \Journal{\ZETF}{34}{247}{1958}
\bibitem{bil78} Bilenky S M and  Pontecorvo B, \Journal{\PRP}{41}{225}{1978}
\bibitem{bil87} Bilenky S M and Petcov S T, \Journal{\RMP}{59}{671}{1987}
\bibitem{kay89} Kayser B, Gibrat-Debu F and Perier F, ``The Physics
of Massive Neutrinos'', {\em World Scientific Lectures Notes in Physics 25},
(Singapore: World Scientific), 1989
\bibitem{dol80} Dolgov A D, Zeldovich Ya B, \Journal{\RMP}{53}{1}{1980}
\bibitem{her93} Herczeg P, {\em Proc. Third Int. Symp. on Weak and 
Electromagnetic Interactions in Nuclei (WEIN-92)}, Dubna, Russia, June 1992,
ed. Vylov Ts, (Singapre: World Scientific) p 262
\bibitem{van93} van der Schaaf A, \Journal{\PPNP}{31}{1}{1993}
\bibitem{dep95} Depommier P and Leroy C, \Journal{\RPP}{58}{61}{1995}
\bibitem{yan79} Yanagida T, {\em Proc. Workshop on Unified Theory
 and Baryon Number in the Universe}, ed. by Sawada and Sugamoto, (KEK, 1979);
 Gell-Mann M, Ramond P and Slansky R, in {\em Supergravity},
 ed. by van Nieuwenhuizen and Freedman, (North-Holland, Amsterdam, 1979)      
\bibitem{bel95} Belesev A I, \Journal{\PLB}{350}{263}{1995}
\bibitem{par96} Particle Data Group, \Journal{\PRD}{54}{1-720}{1996}
\bibitem{ath95} Athanassopoulos C et al, \Journal{\PRL}{75}{2650}{1995}
\bibitem{fuk94} Fukuda Y et al, \Journal{\PLB}{335}{237}{1994} 
\bibitem{bec95} Becker-Szendy R et al, \Journal{\NPBP}{38}{331}{1995} 
\bibitem{goo96} Goodman M, \Journal{\NPBP}{38}{337}{1995}
\bibitem{cle95} Cleveland B T et al, \Journal{\NPBP}{38}{47}{1995}
\bibitem{hir91} Hirata K S et al, \Journal{\PRD}{44}{2241}{1991}
\bibitem{gal95} GALLEX coll, \Journal{\PLB}{357}{237}{1995}
\bibitem{abd94} Abdurashitov J N et al, \Journal{\PLB}{328}{234}{1994}
\bibitem{bah95} Bahcall J N and Pinsonneault M H, 
     \Journal{\RMP}{67}{781}{1995}
\bibitem{mih86} Mikheyev M and Smirnov A, \Journal{\SJNP}{42}{913}{1986};
    Wolfenstein L, \Journal{\PRD}{17}{2369}{1978}, 
    \Journal{\PRD}{20}{2634}{1979} 
\bibitem{oka97} Bilenky S M, Giunti C and Grimus W,
   1998 Eur. Phys. J. C {\bf 1} 247, 1998 Prog. Part. Nucl. Phys. 40;
     Okada N and Yasuda O 1997 Mod. Phys. Lett. {\bf 12} 3669;
    Fogli G L Lisi E Montanino D and Scioscia G 1997 Phys. Rev. D {\bf 56} 4365
\bibitem{sche82} Schechter J and Valle J W F, \Journal{\PRD}{25}{2951}{1982};
  Nieves J F, \Journal{\PLB}{147}{375}{1984}; Takasugi E, 
  \Journal{\PLB}{149}{372}{1984}
\bibitem{hir97t} Hirsch M, Klapdor-Kleingrothaus H V, Kovalenko S G,
    1997 Phys. Lett. B {\bf 398} 311
\bibitem{bil84} Bilenky S M, Nedelcheva N P and Petcov S T, 
    \Journal{\NPB}{247}{61}{1984}; Kayser B, \Journal{\PRD}{30}{1023}{1984}
\bibitem{simf97} \v Simkovic F, Schwieger J, Pantis G, and Faessler A,
    \Journal{\FF}{27}{1275}{1997}
\bibitem{biln97} Bilenky S M, Giunti C, Kim C W and Monteno M,
    1998 Phys. Rev. D {\bf 57} 6981
\bibitem{apo97} Apolonio et al, preprint hep-ex/9711002 and 
    1998  Prog. Part. Nucl. Phys. {\bf 40} 
\bibitem{ach95} Achkar B et al, \Journal{\NPB}{434}{503}{1995}
\bibitem{moh95} Mohapatra R N, Int. Workshop on Neutrinoless Double Beta Decay:
     Double Beta Decay and Related Topics, (Trento, Italy, July 1995)
     ed. H.V. Klapdor-Kleingrothaus and S. Stoica, World Scientific, 
     1996, p. 44
\bibitem{pan96} Hirsch M, Klapdor-Kleingrothaus H V and Panella O,
    \Journal{\PLB}{374}{7}{1996}
\bibitem{pan97} Panella O, Carimalo C, Srivastava Y N and Widom A,
    \Journal{\PRD}{56}{5766}{1997}
\bibitem{wo97} Wodecki A, Kami\'nski W A and Pagerka S,
    \Journal{\PLB}{413}{342}{1997}
\bibitem{bil82} S.M. Bilenky, {\em Introduction to the Physics of Electroweak 
      Interactions} (Pergamon Press Ltd., 1982), p.4.
\bibitem{sim91} F. \v Simkovic, \Journal{\CJF}{41}{1105}{1991};
      F. \v Simkovic and G. Pantis, 1999 to be published in 
      Phys. Atom. Nucl. {\bf 62} 
\bibitem{sim88} \v Simkovic F, \Journal{\CJF}{38}{731}{1988}
\bibitem{bar95} Barbero C, Krmpoti\'c F and Mariano A, 
     \Journal{\PLB}{345}{192}{1995}
\bibitem{civ96} Civitarese O and Suhonen J, \Journal{\NPA}{607}{152}{1996}


\bibitem{moh75} Mohapatra R N and Pati J C, \Journal{\PRD}{11}{2558}{1975};
   Mohapatra R N and Senjanovi\'c G, \Journal{\PRD}{12}{1502}{1975}
\bibitem{moh81} Mohapatra R N and Senjanovi\'c G \Journal{\PRD}{23}{165}{1981}
\bibitem{fri81} Fritzsch H and Minkowski R, \Journal{\PRP}{73}{67}{1981}
\bibitem{doi83} Doi M, Kotani T, Nishiura H and Takasugi E, 
   \Journal{\PTP}{69}{602}{1983}
\bibitem{pan96} Pantis G, \v Simkovic F, Vergados J D and Faessler A, 
   \Journal{\PRC}{53}{695}{1996}
\bibitem{mut95} Muto K, {\em Proc. Int. Workshop on Double Beta Decay and
  Related Topics}, Trento 1995, World Scientific, ed. Klapdor H V and Stoica S
\bibitem{ver90} Vergados J D, \Journal{\NPA}{506}{482}{1990}
\bibitem{pan92} Pantis G, Faessler A, Kami\'nski W A and Vergados J D,
   \Journal{\JPG}{18}{605}{1992}; Pantis G, Vergados J D, 
   \Journal{\PRP}{242}{285}{1994}
\bibitem{krm94} Krmpoti\'c F and Sharma S, \Journal{\NPA}{572}{329}{1994}
\bibitem{bar97} Barbero C, Krmpoti\'c and Tadi\'c, 
    1998 Nucl. Phys. A {\bf 628} 170 
\bibitem{suh91} Suhonen J, Khadkikar S B and Faessler A, 
   \Journal{\NPA}{529}{727}{1991}; \Journal{\PLB}{237}{8}{1990}


\bibitem{gon89} Gonzalez-Garcia and Valle J W F, 
     \Journal{\PLB}{216}{360}{1989}
\bibitem{rom92} Rom\~ao J C, de Campos F and Valle J W F, 
      \Journal{\PLB}{292}{329}{1992} 
\bibitem{aul82} Aulakh C S and Mohapatra R N, \Journal{\PLB}{119}{136}{1982}
\bibitem{bur93} Burgess C P and Cline J M, \Journal{\PLB}{298}{141}{1993};
     \Journal{\PRD}{49}{5925}{1994}
\bibitem{bam95} Bamert P, Burgess C P and Mohapatra R N, 
     \Journal{\NPB}{449}{25}{1995}
\bibitem{car93} Carone C D, \Journal{\PLB}{308}{85}{1993}
\bibitem{bark95} Barbero C, Cline J M, Krmpoti\c F and Tadic D,
     1996 Phys. Lett. B {\bf 371} 78


\bibitem{chan88} Chanowitz M S, \Journal{\ARNPS}{38}{323}{1988}
\bibitem{roy92} Roy D P, \Journal{\PLB}{283}{270}{1992}
\bibitem{bar89} Barger V, Guidice G F and Han T, 
      \Journal{\PRD}{40}{2987}{1989}
\bibitem{zwi83} Zwirner F, \Journal{\PLB}{132}{103}{1983}
\bibitem{wei82} Weinberg S, \Journal{\PRD}{26}{287}{1982}
\bibitem{bar86} Barbieri R and Masiero A, \Journal{\NPB}{267}{679}{1986}
\bibitem{cam91} Campbell B A, Davidson S, Ellis J and Olive K, 
      \Journal{\PLB}{256}{457}{1991}
\bibitem{moh87} Mohapatra R and Valle J W F, \Journal{PLB}{186}{303}{1987}
\bibitem{dre91} Dreiner H and Ross G, \Journal{\NPB}{410}{188}{1994}
\bibitem{val91} Valle J W F, \Journal{\PPNP}{26}{91}{1991}
\bibitem{rom91} Rom\~ao J C, Rius N and Valle J W F, 
      \Journal{\NPB}{363}{369}{1991}
\bibitem{leo86} Leontaris G K, Tamvakis K  and Vergados J D, 
      \Journal{\PLB}{171}{412}{1986}
\bibitem{kos89} Kosmas T S, Leontaris G K and Vergados J D, 
      \Journal{\PLB}{219}{457}{1989}
\bibitem{fae98} Faessler A, Kovalenko S and \v Simkovic F, 
     1998 Phys. Rev. D {\bf 58} 115004
\bibitem{Haber} Haber H E and Kane G L, \Journal{\PRP}{117}{75}{1985}; 
  Gunion J F, Haber H E and Kane G L, \Journal{\NPB}{272}{1}{1986}
\bibitem{pon68} Pontecorvo B, \Journal{\PLB}{26}{630}{1968}


\bibitem{ret95} Retamosa J, Caurier E and Nowacki F, 
     \Journal{\PRC}{51}{371}{1995}; Caurier F, Nowacki F, Poves A and
    Retamosa J, preprint nucl/th9601017
\bibitem{sin88} Sinatkas J, Skouras L D and Vergados J D, 
     \Journal{\PRC}{37}{229}{1988}
\bibitem{gru85} Tomoda T, Faessler A, Schmidt K W and Gr\"ummer F, 
     \Journal{\PLB}{157}{4}{1985}
\bibitem{cau96} Caurier E, Nowacki F, Poves A and Retamosa J, 
      \Journal{\PRL}{77}{1954}{1996}
\bibitem{zha90} Zhao L, Brown B A and Richter W A, 
     \Journal{\PRC}{42}{1120}{1990}
\bibitem{john97} Suhonen J, Divari P C, Skouras L D and Johnstone I P,
      \Journal{\PRC}{55}{714}{1997}
\bibitem{nak96} Nakada H, Sebe T and Muto K, 
      1996 Nucl. Phys. A {\bf 607} 235
\bibitem{pov95} Poves A, Bahukutumbi R P, Langanke K and Vogel P,
      \Journal{\PLB}{361}{1}{1995}
\bibitem{rad95} Raduta A A, Delion D S and Faessler A, 
      \Journal{\PRC}{51}{3008}{1995}
\bibitem{civn95} Civitarese O, Suhonen J and Faessler A, 
     \Journal{\NPA}{591}{195}{1995}; Aunola M, Civitarese O, Kauhanen J 
  and Suhonen J, \Journal{\NPA}{596}{187}{1996}

\bibitem{tom87} Tomoda  T and Faessler A, 
     \Journal{\PLB}{199}{475}{1987}
\bibitem{eng88} Engel J, Vogel P and Zirnbauer M R,
     \Journal{\PRC}{37}{731}{1988}
\bibitem{mut89} Muto K, Bender E and Klapdor H V,
     \Journal{\ZPA}{334}{187}{1989}
\bibitem{mn92} Moeller P and Nix J R, \Journal{\NPA}{536}{20}{1992}. 
\bibitem{ves97} \v Simkovic F, Schwieger J, Veselsk\'y M, Pantis G
      and Faessler A, \Journal{\PLB}{393}{267}{1997}

\bibitem{chi89} Ching C, Ho T and Wu X, \Journal{\PRC}{40}{304}{1989}
\bibitem{sim89} \v Simkovic F, \Journal{\JINR}{39}{21}{1989}
\bibitem{eng92} Engel J, Haxton W C and Vogel P, 
     \Journal{\PRC}{46}{2153}{1992}
\bibitem{sim90} \v Simkovic F and  Gmitro M, 
 {\em Proc. Int. Conf. on Low Energy Weak Interactions (LEWI -90)}, 
 Dubna 1990, p 258
\bibitem{wu91} Wu X, Staudt A, Klapdor H V, Ching C and Ho T,
  \Journal{\PLB}{272}{169}{1991}; 
  Wu X, Staudt A, Kuo T T S and Klapdor H V, \Journal{\PLB}{276}{274}{1992};
 Hirsch M, Wu X R, Klapdor-Kleingrothaus H V,  Ching C and Ho T, 
  \Journal{\ZPA}{345}{163}{1993}
\bibitem{gmi90} Gmitro M and \v Simkovic F, \Journal{\IZV}{54}{1780}{1990}
\bibitem{mut93} Muto K, \Journal{\PRC}{48}{402}{1993}
\bibitem{kad95} Hirsch M, Kadowaki O, Klapdor-Kleingrothaus H V
  and Muto M, \Journal{\ZPA}{352}{33}{1995}
\bibitem{ves98} \v Simkovic and Veselsk\'y, {\em Proc. Int. Workshop MEDEX97},
 Praha, June  1997, \Journal{\CJF}{48}{245}{1998}



\bibitem{ten95} \v Simkovic F, Teneva G, Bobyk A, Cheoun M K,
   Khadkikar S B and Faessler A, \Journal{\PPNP}{32}{329}{1994}; 
   Teneva G, \v Simkovic F, Bobyk A, Cheoun M K, Faessler A 
   and Khadkikar S B,  \Journal{\NPA}{249}{586}{1995}
\bibitem{ber90} Bernabeu J, Desplanques B and Navarro J,
   \Journal{\ZPC}{46}{323}{1990}
\bibitem{rum95} Rumyantsev O A and Urin M G, \Journal{\JETL}{61}{361}{1995}
\bibitem{vlad92} Vladimirov D M and Gaponov Yu V, 
    \Journal{\SJNP}{55}{1010}{1992}
\bibitem{koo97} Koonin S E, Dean D J and Langanke K, 
    \Journal{\PRP}{278}{1}{1997}; Radha P B, Dean D J, Koonin S E, Langanke
    K and Vogel P, \Journal{\PRC}{56}{3079}{1997}


\bibitem{muto97} Muto K, \Journal{\PLB}{391}{243}{1997}
\bibitem{eng97} Engel J, Pittel S, Stoitsov M, Vogel P and Dukelsky J,
     \Journal{\PRC}{55}{1781}{1997}
\bibitem{del97} Delion D S, Dukelsky J and Schuck P, 
     \Journal{\PRC}{55}{2340}{1997}
\bibitem{sam97} Sambataro M and Suhonen J, \Journal{\PRC}{56}{782}{1997}
\bibitem{stoi93} Stoica S and Kami\'nski W A, \Journal{\PRC}{47}{867}{1993}
\bibitem{rad96} Raduta A A and Suhonen J, \Journal{\PRC}{53}{176}{1996} 
\bibitem{eng89} Engel J, Vogel P, Ji X D and Pittel S,
      \Journal{\PLB}{225}{5}{1989}
\bibitem{lyu92} \v Simkovic F, Efimov G V, Ivanov M A and Lyubovitskij V E, 
    \Journal{\ZPA}{341}{193}{1992}

\bibitem{ste98} \v Stekl I, \v Simkovic F, Kovalik A and Brudanin V B, 
    {\em Proc. Int. Workshop MEDEX97},
     Praha, June  1997, \Journal{\CJF}{48}{249}{1998}
\bibitem{gap97} Semenov S V, Gaponov Yu V and Khafizov R U,
    Proc. Int. Workshop on Non-Accelerator New Physics 
    NANP'97 (Dubna, Russia, June 1997), 1998  Yad. Fiz. {\bf 61} 1379;
    Inzhechik L V, Gaponov Yu V and Semenov S V 1998 
    Proc. Int. Workshop on Non-Accelerator New Physics 
    NANP'97 (Dubna, Russia, June 1997), 1998 Yad. Fiz {\bf 61} 1384
\bibitem{kum95} Kume K, {\em Proc. Int. Workshop on Double Beta Decay and
  Related Topics}, Trento 1995, World Scientific, ed. Klapdor H V and Stoica S
\bibitem{bedn97} Bednyakov V A, Brudanin V B, Kovalenko S G and Vylov Ts S,
    \Journal{\MPL}{12}{233}{1997}
\bibitem{hell97} Hellming J and Klapdor-Kleingrothaus H V, 
    \Journal{\ZPA}{359}{351}{1997}; Klapdor-Kleingrothaus H V and
    Hirsch M, \Journal{\ZPA}{359}{361}{1997}


\end{references}
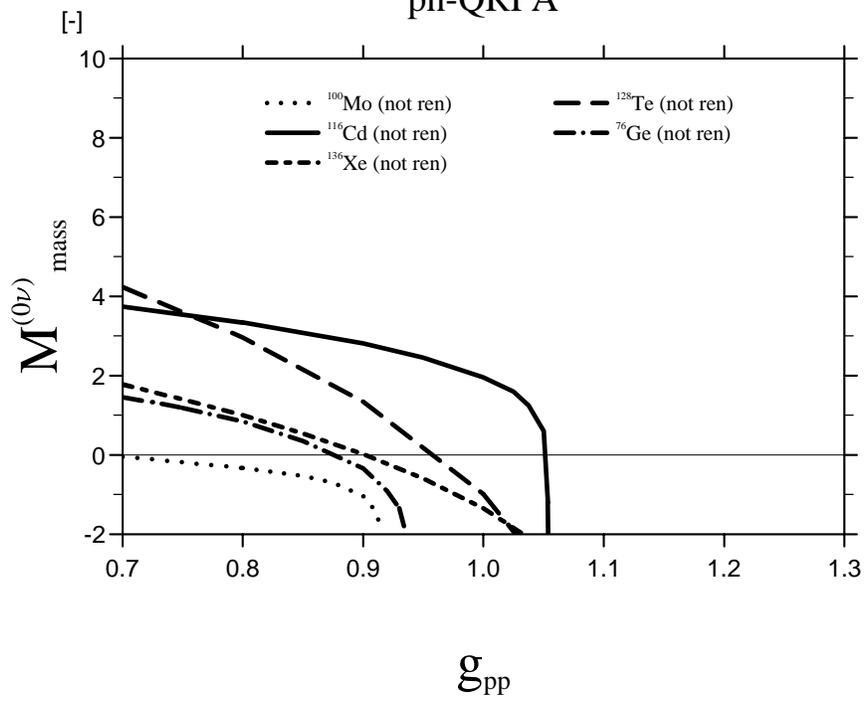
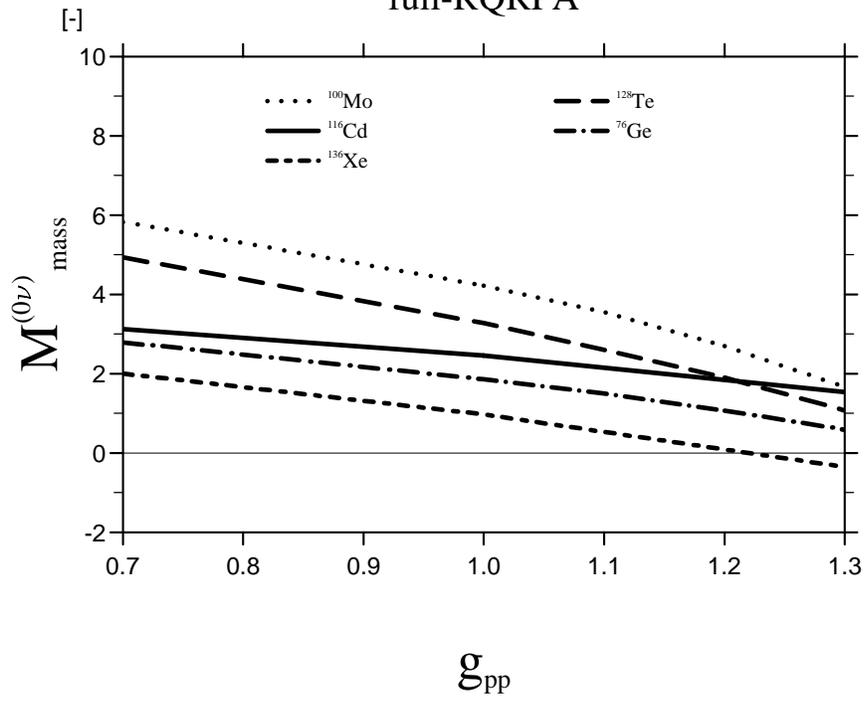
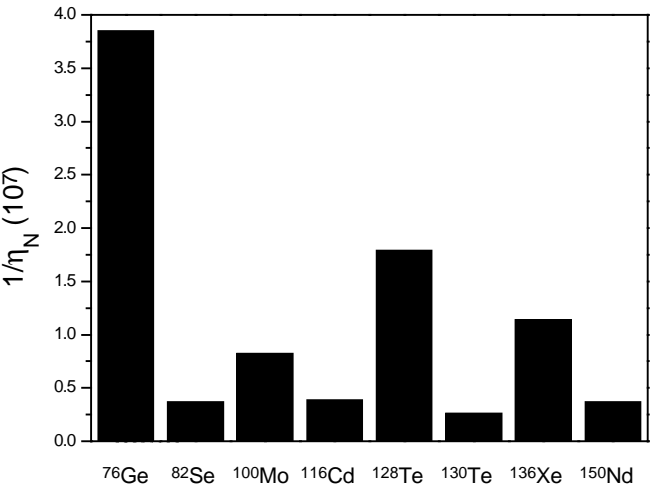
\end{document}